\documentclass[a4paper,12pt]{article}

\usepackage{color}
\usepackage{graphicx}
\usepackage{pictex}
\usepackage{amssymb}
\usepackage{amsbsy}

\newcommand{\newsection}{
\setcounter{equation}{0}
\section}

\def\appendix#1{
  \addtocounter{section}{1}
  \setcounter{equation}{0}
  \renewcommand{\thesection}{\Alph{section}}
  \section*{Appendix \thesection\protect\indent \parbox[t]{11.715cm} {#1}}
  \addcontentsline{toc}{section}{Appendix \thesection\ \ \ #1}
  }

\def\ii{\textcolor{black}{{\bf {$\bullet$}}\ \ }} 
\newcommand{\zag}[1]{\subsection{#1}}
\def\tit{\zag}

\renewcommand{\!}{\negthinspace}

\newcommand{\bbox}[1]{\boldsymbol{#1}}

\def\eop{\vfill\pagebreak}
\def\tre{\textcolor{black}}
\def\tgr{\textcolor{black}}
\def\tma{\textcolor{black}}
\def\tbl{\textcolor{black}}
\def\tcy{\textcolor{black}}
\def\x{\textcolor{magenta}{x}}
\definecolor{gold}{rgb}{1,0.55,0}
\def\tgo{\textcolor{black}}
\definecolor{brown}{rgb}{.63,0.25,0}
\def\tbr{\textcolor{black}}
\def\al{{\textcolor{black}{\alpha}}}
\def\N{\textcolor{black}{N}}

\def\ca{{\textcolor{black}{a}}}

\def\Li{{\rm Li}}

\def\pintab{\int\limits_0^1\hspace{-1.18em}\not\hspace{0.76em}} 

\newcommand{\overset}[2]{\binrel@{#2}%
  \binrel@@{\mathop{\kern\z@#2}\limits^{#1}}}

\newcommand{\tr}[1]{\,{\rm tr}\,#1}

\def\e{{\,\rm e}}
\def\be{\begin{equation}}
\def\ee{\end{equation}}
\def\bea{\begin{eqnarray}}
\def\eea{\end{eqnarray}}
\def\LA{\left\langle}
\def\RA{\right\rangle}
\newcommand{\rf}[1]{(\ref{#1})}
\newcommand{\eq}[1]{Eq.~(\ref{#1})}

\def\bl{{\tgo{\beta}}}

\def\d{{\rm d}}

\def\i{{\rm i}}

\def\s{\sigma}

\def\G{{\cal G}}

\def\Tau{{\cal T}}

\newcommand{\ie}{{i.e.\ }}
\newcommand{\p}{{\prime}}
\newcommand{\ra}{\rightarrow}

\def\lesssim{\mathrel{\mathpalette\fun <}}
\def\gtrsim{\mathrel{\mathpalette\fun >}}
\def\fun#1#2{\lower3.6pt\vbox{\baselineskip0pt\lineskip.9pt
\ialign{$\mathsurround=0pt#1\hfil##\hfil$\crcr#2\crcr\sim\crcr}}}
\def\la{\lesssim}
\def\ga{\gtrsim}

\newcommand{\non}{\nonumber \\*}

\begingroup\makeatletter\ifx\SetFigFont\undefined
\def\x#1#2#3#4#5#6#7\relax{\def\x{#1#2#3#4#5#6}}%
\expandafter\x\fmtname xxxxxx\relax \def\y{splain}%
\ifx\x\y   
\gdef\SetFigFont#1#2#3{%
  \ifnum #1<17\tiny\else \ifnum #1<20\small\else
  \ifnum #1<24\normalsize\else \ifnum #1<29\large\else
  \ifnum #1<34\Large\else \ifnum #1<41\LARGE\else
     \huge\fi\fi\fi\fi\fi\fi
  \csname #3\endcsname}%
\else
\gdef\SetFigFont#1#2#3{\begingroup
  \count@#1\relax \ifnum 25<\count@\count@25\fi
  \def\x{\endgroup\@setsize\SetFigFont{#2pt}}%
  \expandafter\x
    \csname \romannumeral\the\count@ pt\expandafter\endcsname
    \csname @\romannumeral\the\count@ pt\endcsname
  \csname #3\endcsname}%
\fi
\fi\endgroup

\title{{Topics in Cusped/Lightcone Wilson Loops\\[.5cm]} }
\author{
Yuri Makeenko%
\footnote{Also at the Institute for Advanced Cycling,
Blegdamsvej 19, 2100 Copenhagen \O, Denmark}
\\[.2cm]
{\normalsize
{\it Institute of Theoretical and Experimental Physics}} \\
 \normalsize {\it 117218 Moscow, Russia}\\
\normalsize{{makeenko@itep.ru}}
}
\date{\mbox{}}

\begin{document}

\begin{titlepage}

\maketitle

\begin{abstract}
I review several old/new approaches to the string/gauge correspondence
for the  cusped/lightcone Wilson loops. The main attention is payed 
to SYM perturbation theory calculations at two loops and beyond and
to the cusped loop equation. 

These three introductory lectures were given at the
48 Cracow School of Theoretical Physics: ``Aspects of Duality'',
June 13-22, 2008, Zakopane, Poland.
\end{abstract}

\vskip .9cm
\centerline{{\small {\bf Contents}}}
\vspace*{.2cm}

\ii \tma{Lecture 1.}~~\tbr{Pedagogical Introduction} \\
{Wilson loops with cusps, their renormalization, 
relation to twist-two operators,
the role in string/gauge correspondence, minimal surface in 
$AdS_5\otimes S^5$ for cusped loops};  

\ii \tma{Lecture 2.}~~\tbr{Perturbation Theory: 
two loops and beyond}
\\
{exact sum of ladders, explicit two loops and the anomaly terms, 
results in the double logarithmic approximation, problems with planar QFT}; 

\ii \tma{Lecture 3.}~~\tbr{Cusped Loop Equation}
\\
modern formulation of the loop equation, SUSY extension, UV regularization,
specifics of cusped loops, cusp anomalous dimension from the 
loop equation.

For completeness of these lecture notes I added three appendices with
some detail not given in the lectures but which might be useful
for the reader.

\end{titlepage}
\setcounter{page}{2}

\newsection*{Lecture 1.~~Pedagogical Introduction}
\setcounter{section}{1}

I review in this lecture Wilson loops with cusps, their renormalization and 
the relation to twist-two operators. Then I discuss 
the role played by the cusped Wilson loops in the string/gauge correspondence 
and describe a proper minimal surface in 
$AdS_5\otimes S^5$ associated with cusped loops.  

\zag{Wilson loops}

Wilson loops play a crucial role in modern formulations of gauge theories
since the work by Wilson (1975).

The construction is based on a non-Abelian phase factor
\be
U(\tgr{C})= \bbox{P} \e^{\i \tgo{g} 
\int_{\tgr{C}} A_\mu(x) \d x^\mu} \stackrel{{\rm def}}=
\prod_{x\in \tgr{C}} \left( 1+ \i \tgo{g} A_\mu(x) \d x^\mu \right)
\ee
which is nothing but a parallel transporter in an external 
 non-Abelian Yang--Mills field $ A_\mu(x)$.
The trace over matrix indices 
$
\tr U(\tgr{C}) 
$
is gauge-invariant for closed $\tgr{C}$.

The Wilson loop vacuum expectation value 
(or the average in Euclidean formulation) is defined by
\be
W(\tgr{C})=Z^{-1} \int {\cal D}A_\mu \,{\cal D}\bar \psi {\cal D}\psi \cdots \,
\e^{\i S} \frac 1{\N} \tr U(\tgr{C})\,, 
\ee
where the path integration goes over Yang-Mills and quark fields.

The importance of the Wilson loops in QCD is because\\
\ii observables are expressed via sum-over-path of $W(\tgr{C})$,\\
\ii dynamics is entirely reformulated via $W(\tgr{C})$.

These statements hold strictly speaking only at large $\N$, while at finite
$\N$ correlators of several Wilson loops appear which factorize in the
large-$\N$ limit.  
Wilson  loops $W(\tgr{C})$ obey the \tbr{loop equation} which is 
a closed equation on loop space at large $\N$
(see the book by Y.~M. (2002) for more detail on these issues).

It is important that 
typical loops which are essential in the sum-over-path are {\it cusped}.
The properties of the Wilson loops of
this kind differ from those for smooth loops, e.g.\ a circular loop.

\tit{Renormalization of smooth Wilson loops}

Renormalization properties of
{\it smooth}\/ Wilson loops are studied by {Gervais, Neveu (1980)},
{Polyakov (1980)}, {Vergeles, Dotsenko (1980)}. 
They become finite after  the charge renormalization:
\be
W(\tgo g; \tgr C)= \e^{-{\rm const.} \,L(C)/\tcy{a}}\;
W_{\rm R}(\tgo g_{\rm R} ; \tgr C)\,,
\label{per}
\ee
where $W_{\rm R}$ is  \tma{finite} after the {charge renormalization} 
$\tgo{g}\Longrightarrow\tgo g_{\rm R}$
and $\tcy{a}$ is a certain (gauge-invariant) UV cutoff.

The exponential perimeter factor in \eq{per} is associated with 
 the renormalization of the mass of a heavy test
particle propagating along the loop.
It does {not} emerge in {dimensional regularization}.

\tit{Renormalization of cusped Wilson loops}

\begin{figure}[h]
\vspace*{3mm}
\centering{\includegraphics{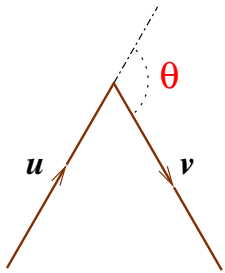} 
}
\caption[cusped Wilson loop]  
{Segment of a closed loop near the cusp.
The cusp angle $\tre\theta$ is formed by the vectors $u$ and $v$:
$\cosh \tre \theta=\frac{u\cdot v}{\sqrt{u^2}\sqrt{v^2}}$.}
   \label{fi:cusped}
\end{figure}

An additional {logarithmic} divergence appears for cusped loops 
as was first discovered by {Polyakov (1980)}. 
A cusped Wilson loop is depicted in Fig.~\ref{fi:cusped}.

The cusped Wilson loops are {multiplicatively renormalizable} as was
shown by {Brandt, Neri, Sato (1981)}:
\be
W(\tgo g; \tgr \Gamma)= 
Z(\tgo g, ; \tre{\theta})
W_{\rm R}(\tgo g_{\rm R}; \tgr \Gamma)\,,
\label{clr}
\ee
where (the {divergent} factor of) $Z(\tgo{g};\tre\theta)$ 
depends on the cusp angle $\tre\theta$.

Equation~\rf{clr} is true only when the contour 
 $\tgr{\Gamma}$ has no light-cone segments.
The peculiarities of the renormalization of light-cone Wilson loops are
described below.

\tit{Cusp anomalous dimension}

The cusp anomalous dimension is defined by the formula
\be
\gamma_{\rm cusp}\left(\tgo{g};\tre \theta\right)=
-\ca \frac{\d}{\d \ca} \ln Z(\tgo g ;\tre \theta)\,,
\label{defgamma}
\ee
where $Z$ is the renormalizing factor in \eq{clr}.

The cusp anomalous dimension depends in general both on the coupling 
constant $g$ and on the cusp angle $\tre \theta$.  
A very important observation by {Korchemsky, Radyushkin (1987)} is 
that in {the limit of large} $\tre \theta$ it is linear in $\tre \theta$:
\be
\gamma_{\rm cusp}
\left(\tgo g ;\tre\theta\right) 
\stackrel{\tre \theta \rightarrow\infty}\rightarrow\frac{\tre\theta}{2}
f(\tgo g) \,. 
\ee
The same function $f(\tgo g) $ {appears in the anomalous dimensions of 
twist-two conformal operators with large spin}.

\zag{Conformal operators of twist two}

Anomalous dimensions of twist-two operators of the type
\be
O_{\tbr{J}}^{(F)}=\frac 1N \tr{} F_{\mu \tbr{\cdot} } 
\left(\nabla_{\tbr\cdot} \right)^{\tbr J-2}  F_{\mu \tbr{\cdot} } 
\label{OF}
\ee
\be
O_{\tbr{J}}^{(\Psi)}=\bar\Psi \gamma_ {\tbr{\cdot}}
\left(\nabla_{\tbr\cdot} \right)^{\tbr J-1}  \Psi \Big.
\label{Opsi}
\ee
{with Lorentz spin} $\tbr{J}$  are measurable in deep inelastic scattering.
The operators in \eq{OF} are constructed from gauge field and those in \eq{Opsi}
are constructed from quarks.

In ${\cal N}=4$ supersymmetric Yang--Mills (SYM) there are also
analogous operators
\be
O_{\tbr{J}}^{(\Phi)}=\frac 1N \tr{} \Phi
\left(\nabla_{\tbr\cdot} \right)^{\tbr J}  \Phi 
\label{Ophi}
\ee
constructed from scalars. 

The following notation is used in Eqs.~\rf{OF}, \rf{Opsi} and \rf{Ophi}:
\be
\qquad \nabla_{\tbr \cdot} 
\equiv \nabla_{\tbr \mu} \xi_{\tbr \mu} \qquad \xi^2=0 \,.
\ee
This multiplication by a light-like vector $\xi$ provides
 symmetrization and subtraction of traces, as is needed for a representation 
of the Lorentz group.

What is depicted in Eqs.~\rf{OF}, \rf{Opsi} and \rf{Ophi} by
$\left(\nabla_{\tbr\cdot} \right)^{\tbr J} $ is in fact a
polynomial in
$\stackrel{\leftarrow} \nabla_{\tbr \cdot}$ and 
$\stackrel{\rightarrow} \nabla_{\tbr \cdot}$, the covariant derivatives acting 
on the left and on the right. At the one-loop level it is 
a Gegenbauer polynomial dictated by conformal invariance as was shown by 
{Brodsky, Frishman, Lepage, Sachrajda (1980)}, {Y.~M. (1981)},
{Ohrndorf (1982)}.
These conformal operators are multiplicatively renormalizable at one loop
and form a convenient basis for two-loop computations.

\tit{Relation between the anomalous dimensions}

\begin{figure}[h]
\vspace*{3mm}
\centering{\includegraphics{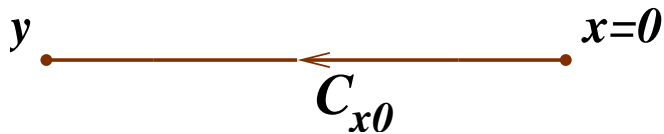} 
}
\caption[straight Wilson loop]  
{Straight Wilson loop from $0$ to $y$.}
   \label{fi:straight}
\end{figure}

The relation between twist-two operators and cusped Wilson loops
can be understood by considering an {open Wilson loop} with matter
fields attached at the ends:
\be
O(\tgr C_{y0})=\bar\psi(y) {\bbox P}\e^{\i \tgo  g
\int_0^y \d \xi^\mu A_\mu } \psi(0)\,.
\ee
The case when this open loop is a straight line from $0$ to $y$
is depicted in Fig.~\ref{fi:straight}.

\begin{figure}[h]
\vspace*{3mm}
\centering{\includegraphics{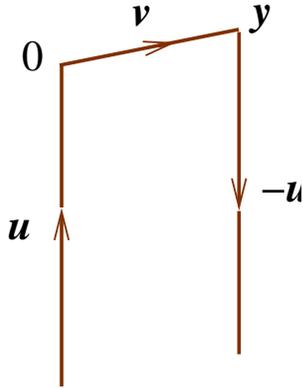} 
}
\caption[pishaped Wilson loop]  
{$\tgr\Pi$-shaped Wilson loop.}
   \label{fi:pishaped}
\end{figure}

The standard triangular diagrams, which give the anomalous dimension
like in Gross, Wilczek (1973), come from the formula
\be
\LA \psi(\infty,\vec y) O(\tgr C_{y0})\bar\psi(\infty,\vec 0)  \RA
\propto W(\tgr\Pi)
\label{Pi}
\ee
as mass of matter fields $\ra \infty$. The Wilson loop which emerges
on the right-hand side is $\Pi$-shaped as is depicted in Fig.~\ref{fi:pishaped}.
The vertical lines represent propagation (in time) of static quarks, 
sitting at $\vec 0$ and $\vec y$, which are connected by the 
straight line associated with the open Wilson loop.

To derive \eq{Pi}, 
{remember that} the propagator in an external field $A_\mu$ is
\be
\LA \psi_i(x) \bar\psi_j(y) \RA_\psi 
\stackrel{\tma{\rm large}~ \N}= \sum _{\tgr C_{yx}} 
\left[\e^{\i \tgo{g} \int_{\tgr C_{yx}} \d \xi^\mu A_\mu }\right]_{ij}
\stackrel{\tma{\rm mass}\rightarrow\infty}
\propto 
\left[
\e^{\i\tgo{g} \int_{\tgr C^{\tre{\rm (min)}}_{yx}}\d \xi^\mu A_\mu }\right]_{ij}
\ee
and thus the {straight vertical} lines appear in $\tgr\Pi$.

The central segment of $\tgr{\Pi}$ is near the light-cone
to suppress the contribution from operators of twists higher than two.
$\tgr{\Pi}$ has two cusps with $\tre\theta\rightarrow\infty$.

This is how the {light-cone} Wilson loop is related to
the conformal operators of twist two.

\tit{Light-cone Wilson loops}

For the $\tgr{\Pi}$-shaped loop with 
{1 light-cone} segment as in Fig.~\ref{fi:pishaped}, we have  
\be
W\left( \tgr{\Pi}  \right) = \e^{-\frac12 f(\lambda) \ln^2 \frac T\ca +
{\rm const.}(\lambda)\ln \frac T\ca +{\rm finite}(\lambda)} 
\label{Km}
\ee
with the same $f(\lambda)$ as before, as was shown by 
{Korchemsky, Marchesini (1993)}. Here
$v^\mu$ is along the light cone ($v^2=0$) and $y_\mu=v_\mu T$. 

A very closely related (and more simple!) object, 
proposed by {Alday, Maldacena (2007)}, is
a  $\tgr{\Gamma}$-shaped loop which is formed by 
2 light-cone segments: 
\be
W\left( \tgr{\Gamma}  \right) = \e^{-\frac12 f(\lambda) 
\ln \frac T\ca \ln \frac S\ca +
{g}(\lambda)(\ln \frac T\ca +\ln\frac S\ca) +{\rm finite1}(\lambda)} \,.
\ee
Now both $v^\mu$ and $u^\mu $ are along the light cones ($v^2=0$, $u^2=0$) 
and $y_\mu=v_\mu T$, $x_\mu=u_\mu S$. 

Most probably it gives the same $f(\lambda)$ but this is not yet rigorously proved.

\tit{SYM Wilson loops} 

{An extension of Wilson loops to} ${\cal N}=4$ supersymmetric Yang--Mills (SYM) 
was given by {Maldacena (1998)}: 
\be
W_{\tre{\rm SYM}}(\tgr C)= \left\langle \frac 1{\N} \tr 
\,{\bbox P} 
\e^{\i \tgo g 
\oint_{\tgr C} \d \sigma \left(\dot\xi^\mu A_\mu
+|\dot \xi | n^i \Phi_i\right)}\right\rangle
\label{WSYM}
\ee
with unit vector $n^i\in S^5$  ($n^2=1$) and 6 scalars $\Phi_i$ 
{($i=1,\cdots,6$)}. In Minkowski space there is no relative
 $\i $ between the two terms in the exponent in \eq{WSYM}, which is
present in Euclidean space. 

Under a supersymmetry transformation 
\be
\delta A_\mu = \bar \Psi \Gamma_\mu \zeta\,, \qquad 
\delta \Phi_i =\bar \Psi \Gamma_i \zeta\,, 
\label{Str}
\ee
where an infinitesimal parameter $\zeta$ is a 10d Majorana--Weyl spinor
and $(\Gamma_\mu ,\Gamma_i)$ are 10d gamma matrices,
the SYM Wilson loop~\rf{WSYM} remains unchanged 
if
\be
 \left(\Gamma_\mu \dot\xi^\mu + \Gamma_i |\dot \xi| n^i \right)\zeta =0\,.
\label{nilp}
\ee
Noting that the combination of gamma matrices in the brackets is nilpotent
for timelike $\dot \xi$, \eq{nilp} is satisfied when a half components of
$\zeta$ vanish. An example of such a BPS state possessing a half of 
supersymmetries is 
the SYM Wilson loop for a straight line inside the light-cone, for which 
we have
\be
W_{\tre{\rm SYM}}(|)=1 \,.
\label{=1}
\ee

An adjoint Wilson loop is related to the fundamental-representation one
by the formula
\be
 \tr_{\!\!A}\,U= | \tr U|^2 -1 \,.
\ee
Due to factorization at large $\N$ we have
\be
\LA \frac1{\N^2}\tr_{\!\!A}\,U(C)\RA = \LA \frac1\N\tr U(C)\RA^2 \,,
\ee
where the {adjoint} Wilson loop is on the left-hand side and 
the (square of the) {fundamental} one is on the right-hand side.

The same results as mentioned above for QCD hold for SYM Wilson loops 
and there are some more. In particular, the perimeter factor in \eq{per}
is missing for SYM Wilson loops owing to the cancellation between
gauge fields and scalars.

\zag{Motivation (since 2002)\label{GKP}}

A remarkable 
 prediction for the anomalous dimension of twist-two operators 
with large (Lorentz) spin $J$,
based on the AdS/CFT correspondence, was made by 
{Gubser, Klebanov, Polyakov (2002)}. It states that
\be
\Delta-\tbr{J}-2 = f(\lambda) \ln \tbr{J} \qquad \hbox{large }\tbr{J}
\label{(1)}
\ee
with
\be
f(\lambda)= \frac{\sqrt{\lambda}}\pi   
~~\qquad \hbox{large }\lambda=\tgo{g}^2_{\rm YM} \N \,.
\label{(2)}
\ee
It stems from 
the spectrum of closed folded string which is rotating in $AdS_5$. 

The 
same result holds for the {cusp anomalous dimension} at large $\tre\theta$
in the supergravity approximation to the AdS/CFT 
correspondence as
is demonstrated by {Kruczenski (2002)}, {Y.~M. (2003)}
and reviewed in the lectures by L.~Alday at this School.
For completeness of these lectures I describe the proper minimal surface 
in $AdS_5 \otimes S^5$ in Appendix~A.

Equations~\rf{(1)}, \rf{(2)} have been remarkable reproduced recently 
from the spin chain $S$-matrix by Staudacher (2005), {Eden, Staudacher (2006), 
Beisert, Eden, Staudacher (2007)}. Many more results are
obtained along this line as is reviewed in the lectures by M.~Staudacher
at this School.

Remarkably, the
same function $f(\lambda)$ appears in MHV gluon amplitudes for SYM
as conjectured by {Bern, Dixon, Smirnov (2005)} on the basis of a few
lower orders of SYM perturbation theory and further elaborated by
Bern, Czakon, Dixon, Kosower, Smirnov (2007). This subject is reviewed in
the lectures by R.~Roiban at this School.
The BDS amplitude
is reproduced for large $\lambda$ from 
the AdS/CFT correspondence by {Alday, Maldacena (2007)} and is reviewed
in the lectures by L.~Alday at this School.

It is a challenging problem to obtain $\sqrt{\lambda}$ for the cusp anomalous
dimension at large  
$\lambda$ in SYM perturbation theory. I shall describe some steps along this 
line in my second and third lectures. 

\tit{AdS/CFT for Wilson loops\label{1.10}}

\begin{figure}[h]
\vspace*{3mm}
\centering{\includegraphics{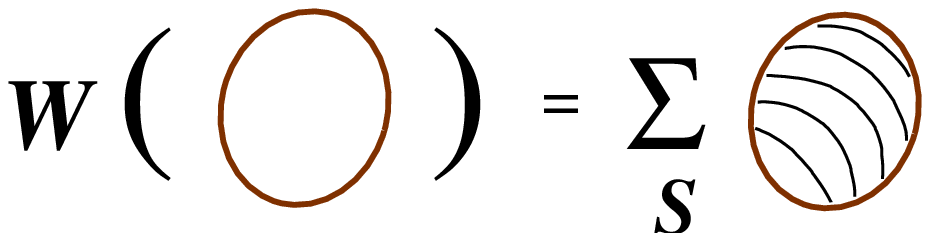} 
}
\caption[AdS/CFT Wilson loop]  
{Open-string/Wilson-loop correspondence.}
\label{fi:AdS/CFT}
\end{figure}

The formulation of the AdS/CFT correspondence between Wilson loops and
open IIB strings in the $AdS_5\otimes S^5$ background was given by 
{Maldacena (1998)}, {Rey, Yee (1998)}.
The statement is that the SYM Wilson loop equals the sum over 
open surfaces bounded by the contour $C$:
\be
W_{\tma{\rm SYM}}(\tgr C)= \sum \limits _{S:\partial S=\tgr C}
\e^{\i A_{\tma{IIB~{\rm on} ~ AdS_5\otimes S^5}}}\,.
\label{AdS/CFT}
\ee
Here
\be
\tgr  C = \left(x^\mu(\sigma), 
\int^\sigma \d \sigma\, |\dot x(\sigma)|n^i(\sigma)\right)
\label{9-dim}
\ee 
is a (9-dimensional loop) in the  boundary of $AdS_5\otimes S^5$,
e.g.\ for $n^i=(1,0,0,0,0,0)$ only a 4d contour $x^\mu(\sigma)$ remains.
Equation~\rf{AdS/CFT} is graphically represented in Fig.~\ref{fi:AdS/CFT}.

For a circular loop there is a remarkable perfect agreement between the 
AdS supergravity calculation by {Berenstein, Corrado, Fischler, Maldacena 
(1998)}, {Drukker, Gross, Ooguri (1999)}
and the {CFT} SYM calculation by {Erickson, Semenoff, Zarembo (2000)},
{Drukker, Gross (2001)}.

\begin{figure}[h]
\centering{\includegraphics{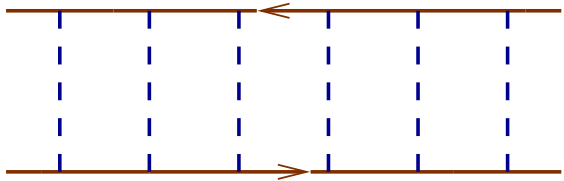} 
}
\caption[ladders]  
{Ladder diagrams for a rectangular Wilson loop.}
\label{fi:ladd}
\end{figure}

However, the situation is not as good for 
a rectangular loop (or antiparallel lines), when the 
minimal surface in $AdS_5\otimes S^5$ was found by {Maldacena (1998)},
{Rey, Yee (1998)}, and the summation of ladder diagrams of the type
depicted in Fig.~\ref{fi:ladd} was performed by 
{Erickson, Semenoff, Szabo, Zarembo (1999)}, 
{Erickson, Semenoff, Zarembo (2000)}.

\noindent The results are: \\
\begin{minipage}[t]{0.5\textwidth}
\vfill
\tre{AdS:}
$$V(\tbl R)=-\frac{4\pi^2 \sqrt{2\lambda}}{\Gamma^4(1/4)\tbl R}$$
\end{minipage}
\hfill 
\begin{minipage}[t]{0.5\textwidth}
\vfill
\tre{SYM:}\be V(\tbl R)=-\frac{\sqrt{\lambda}}{\pi \tbl R}\,. \ee
\end{minipage}
\vspace*{.2mm}

The coefficients in these two results obviously do not agree.
The discrepancy is customly attributed to interaction diagrams...
But a remark is that the SYM coefficient 
is what is needed for the cusp anomalous dimension at large $\lambda$.

\newsection{Perturbation Theory: two loops and beyond}

I describe in this lecture how to sum up ladder diagrams of
perturbation theory and explicitly consider the two-loop order where
a cancellation of interaction diagrams is not complete resulting
in an anomaly term. I analyze
light-cone Wilson loops in the double logarithmic approximation  
which gives a hint on higher-order anomaly terms and reveals 
problems with planar QFT in describing the results.

\tit{One-loop perturbation theory}

Diagrams of perturbation theory for SYM Wilson loops can be constructed 
by expanding \eq{WSYM} in the coupling constant $\lambda$.

To the order $\lambda$ (one-loop order) we have explicitly
\be
W(\tgr \Gamma) = 1- \frac{\lambda}{2} \int\limits_{-\infty}^{+\infty}
\d \sigma_1 \int\limits_{-\infty}^{+\infty}
\d \sigma_2 \,\Big[ \dot x^\mu (\sigma_1) \dot x_\mu(\sigma_2)
-|\dot x (\sigma_1)|| \dot x(\sigma_2)| \Big] 
\tbl D\left(x(\sigma_1)-x(\sigma_2)\right),
\label{ol1}
\ee
where
\be
\tbl D(x)=-\frac{\Gamma\left( d/2-1\right)}{4\pi^{d/2}} [-x^2]^{1-d/2}
\ee
is the (scalar) propagator in $d$-dimensions.

\begin{figure}[h]
\hspace*{5mm}
{\includegraphics[width=14cm]{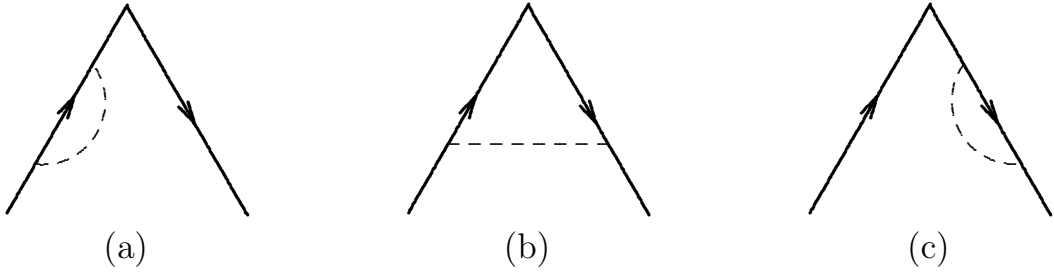} 
}
\caption[one-loop perturbation theory]  
{Diagrams of perturbation theory to order $\lambda$.
Diagrams (a) and (c) {vanish} ({gluons} are cancelled by {scalars}).}
\label{fi:1-loop}
\end{figure}

The double integral 
on the right-hand side of \eq{ol1} can be represented as the sum of
three diagrams in Fig.~\ref{fi:1-loop}, where dashed lines correspond
to either gluon or scalar propagators.
Actually, the diagrams (a) and (c) {vanish} because {gluons} 
are cancelled by {scalars} (like in \eq{=1}).

For the only nonvanishing diagram in Fig.~\ref{fi:1-loop}(b), 
we have\\
\begin{minipage}[t]{0.1\textwidth}
\vfill
\includegraphics[width=4cm,height=5cm]{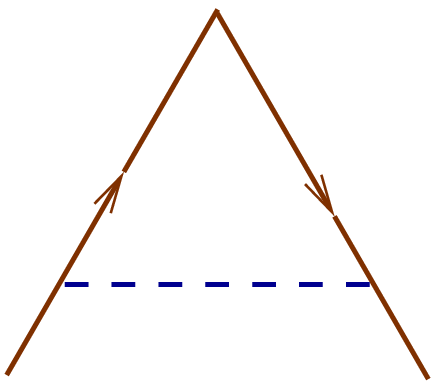}  
\end{minipage}
\hfill 
\begin{minipage}[t]{0.9\textwidth}
\vfill
\bea
W(\tgr \Gamma)&=& 1-
\frac{\lambda}{4 \pi^2} (\cosh \tre\theta-1)
\int \d s \int \d t\,
\frac{1}{s^2 + 2 st\cosh  \tre\theta + t^2} \non
&&\hspace*{1cm}=1- \frac{\lambda}{4 \pi^2} 
\frac{\cosh \tre \theta-1}{\sinh \tre \theta}\tre\theta \, \ln \frac L\ca
\non
&&\hspace*{2cm}\stackrel{{\rm large~} \tre\theta}\ra
1- \frac{\lambda}{4 \pi^2}\tre \theta \, \ln \frac L\ca
\label{1ll}
\eea
\be
\hbox{\tma{which yields}}\qquad
\tbl\Longrightarrow \qquad f(\lambda)=\frac{\lambda}{2 \pi^2} \,.
\ee
\end{minipage}
\vspace*{2mm}

Notice that, in contrast to QCD, in SYM  there is
no mass-renormalization term $-\lambda/4\pi\ca$. 

An exact formula for the order $\lambda$ result is
\begin{eqnarray}
\lefteqn{\hspace*{-.8cm}
W(S,T;a,b)= 1-
\frac{\lambda}{4 \pi^2} (\cosh \tre\theta-1)
\int_a^S \d s \int_b^{ T} \d t\,
\frac{1}{s^2 + 2 st\cosh \tre\theta + t^2}} 
\nonumber \\*
&=&1-\frac{\lambda}{8\pi^2} \frac{\cosh  \tre\theta-1}{\sinh  \tre\theta}
\left( \Li_2(-\frac T{S} \e^{\tre\theta})-\Li_2(-\frac T{ S} \e^{- \tre\theta})
-\Li_2(-\frac T{a} e^{ \tre\theta}) \right.\nonumber \\*
&&\hspace*{0cm} \left.+
\Li_2(-\frac T{ a} e^{- \tre\theta})-\Li_2(-\frac b{S}
\e^{ \tre\theta})+\Li_2(-\frac b{S} \e^{- \tre\theta})
+\Li_2(-\frac b{a} \e^{ \tre\theta})-
\Li_2(-\frac b{a} \e^{- \tre\theta}) \right)
\label{li's}
\end{eqnarray}
where we cut the integrals by $S,T$ from above and by $a,b$ from below.

In \eq{li's} $\Li_2$ is {Euler's dilogarithm}
\be
\Li_2(z)= \sum_{n=1}^\infty \frac{z^n}{n^2}=-\int_0^z \frac{d x}{x}\,
\ln{\left(1-x\right)}
\ee
which obeys the relation
\be
\Li_2\left(-e^\Omega\right)+\Li_2\left(-e^{-\Omega}\right)=
-\frac 12 \ln^2 \Omega -\frac{\pi^2}6 \,.
\ee
It is used to extract the double logarithms.

\tit{Double-logarithmic approximation}

The final result for large $\theta$ in \eq{1ll} can be extracted 
without an exact computation
using a double logarithmic approximation (DLA)
quite similar to the one for Sudakov's form-factor.

The one-loop integral 
\be
W(S ,T;a,b)= 1-
2 \bl (\cosh \tre\theta-1)
\int_a^S \d s \int_b^T \d t\,
\frac{1}{s^2 + 2 st\cosh  \tre\theta + t^2} \,,
\label{11ll}
\ee
where we have introduced
\be
\bl = \frac{\lambda}{8 \pi^2}\,,
\label{bl}
\ee
has a  {\it double-logarithmic}\/ region of integration: 
\be
t\e^{-\tre\theta} \lesssim s \lesssim t\e^{\tre\theta}\qquad \hbox{or}\qquad
s\e^{-\tre\theta} \lesssim t \lesssim s\e^{\tre\theta} \,.
\ee
As a consequence of this fact, we write \eq{11ll} for $\tre\theta \gg1$
in DLA as 
\be
{W(S ,T;a,b)\stackrel{\rm DLA}=1-  \bl
\int\limits^T_{ b} 
\frac{\d t}t \int\limits_{{\rm max}\{a,t\e^{-\tre\theta}\}}^{{\rm min}\{
S,t\e^{\tre\theta}\}}
\frac{\d s}s} \,.
\ee

Several limits are now possible. Let 
\be
1\ll\tre\theta \la \ln \frac Tb \ll \ln \frac Sa \,.
\ee
Then
\be
W(S ,T;a,b)\Longrightarrow 1-  2 \bl \tre\theta
 \ln \frac{T}{b} 
\ee
reproducing the above result~\rf{1ll}. 

Alternatively, if 
\be
\tre\theta \gg \ln \frac Tb,\;  \ln \frac Sa \,,
\ee
we obtain
\be
W(S ,T;a,b)\Longrightarrow 1-   \bl \ln \frac{T}{b} \ln \frac{S}{a} 
\qquad \fbox{very large
$\tre\theta$ }
\ee
reproducing the result for 2 light-cone segments.

Some more results on the double logarithms are described in Appendix~B.

\tit{Sum of ladder diagrams}

\begin{figure}[h]
\centering{\includegraphics[width=5cm,height=5cm]{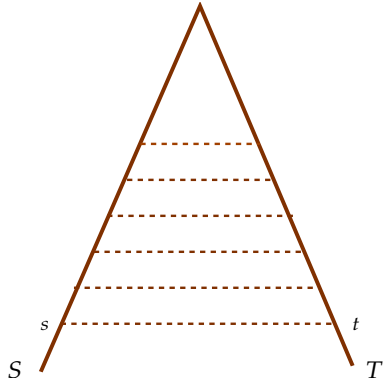}  
}
\caption[ladder diagram]  
{Typical ladder diagram for a cusped Wilson loop.}
\label{fi:ladder}
\end{figure}

As was explained in the first lecture, we are motivated to consider
ladder diagrams of the type depicted in Fig.~\ref{fi:ladder}, 
which is a simplest class of diagrams of planar QFT.
 
The ladder diagrams can be summed using a 
Bethe--Salpeter equation
\be
\G(S,T)= 1-\frac{\lambda(\cosh \tre\theta-1)}{4\pi^2}
\int_a^S \d s \int_b^T \d t 
\frac{\G(s,t)}{s^2+2 st \cosh \tre\theta +t^2} \,.
\ee

Several light-cone limits are again possible:\\
{\bf 1 light-cone limit}: $\tre\theta\to\infty$ with fixed 
$T_{\rm l.c.}=2T\e^{\tre\theta}$ resulting in 
\begin{equation}
\G\left(S,T;a,b\right)=1-\bl \int_a^S ds \int_b^{ T} dt\,
\frac{\G\left(s,t;a,b\right)}{\al s^2 + st} \,,
\label{nonladder2}
\end{equation}
where
\be
\al=\frac{u^2}{2 u\cdot v} 
\ee
and $\bl$ is given by \eq{bl}
{(remember that $v^2=0$ for the light-cone direction).}\\
{\bf 2 light-cone limit}: 
for $\al=0$ when additionally $u^2=0$, the cusped loop has 2 light-cone
segments. 

\tit{The ladder equation}

Differentiating Eq.~(\ref{nonladder2}) we obtain
\begin{equation}
S\frac{\partial}{\partial S}\,T\frac{\partial}{\partial T}\,
\G\left(S,T;a,b\right)=
-\frac{\bl}{1+\al S/T}~
\G\left(S,T;a,b\right)
\label{edifladder2}
\end{equation}
and analogously
\be
a\frac{\partial}{\partial a}\,b\frac{\partial}{\partial b}\,
\G\left(S,T;a,b\right)=
-\frac{\bl}{1+\al a/b}~
\G\left(S,T;a,b\right)
\label{edifladder1}
\ee
with the {boundary conditions}
\begin{equation}
\G(a,T;a,b)=\G(S,b;a,b)=1\,.
\label{bc}
\end{equation}

To separate variables, it is convenient to introduce the new variables 
\be
X =\ln \frac Sa-\ln \frac Tb \,, \qquad 
Y=\ln \frac Sa+\ln \frac Tb \,.
\ee
Then Eqs.~\rf{edifladder2} and \rf{edifladder1} 
can be rewritten as
\be
\left(\frac{\partial^2}{\partial X^2}-\frac{\partial^2}{\partial Y^2}
\right)\G= \frac{\bl}{1+\al \frac {a}b\e^X}\,\G ~\stackrel{\al S\ll T}{=}
\bl\,\G \,.
\label{ladde}
\ee
It is similar to the equation by
{Erickson, Semenoff, Szabo, Zarembo (1999)}
but with {different boundary conditions}.

\tit{Exact solution for ladders ($\tbr{\alpha}= 0$)}

A solution to \eq{ladde} for $\al=0$ is the {Bessel function}
\be
\G_{\al=0}\left(S,T;a,b\right)= J_0 \left( 2\sqrt{
\tgo\beta \ln \frac Sa \ln \frac Tb}\right)
\ee
which obviously obeys the boundary condition~\rf{bc}.

This can be easily shown by an {iterative solution} of
\be
\G_{\al=0}\left(S,T;a,b\right)=1-\bl \int_a^S \frac {\d s}s 
\int_b^{ T} \frac{\d t}t\,
\G_{\al=0}\left(s,t;a,b\right), 
\ee
where the integrals over $s$ and $t$ {decouple} and 
both are {logarithmic}:
\be
\G_{\al=0}\left(S,T;a,b\right)=\sum_{n=0}^\infty (-\tgo \beta)^n
\frac{\left(\ln \frac Sa\right)^n}{n!}\frac{\left( \ln \frac Tb\right)^n}{n!}
=J_0 \left( 2\sqrt{
\tgo\beta \ln \frac Sa \ln \frac Tb}\right).
\ee

{Asymptotically} we have
\be
J_k(z)\sim \cos z \qquad\quad\hbox{large\ }\;z
\ee
which is {\em not}\/ of the type expected in \eq{Km} from {renormalization}.

\tit{Exact solution for ladders  ($\tbr{\alpha}\neq 0$)}

An exact solution to \eq{ladde} for $\tbr{\alpha}\neq 0$ is found
by {Y.~M., Olesen, Semenoff (2006)}. 

Let us consider the ansatz
\begin{equation}
\G\left(S,T;a,b\right) =\oint\limits_C \frac{\d\omega}{2\pi i \omega}
\left(\frac S{a} \right)^{\sqrt{\bl}\omega }
\left(\frac T{b} \right)^{-\sqrt{\bl}\omega^{-1} }
\,F\left(-\omega, \al \frac ab \right)
F\left(\omega, \al \frac ST \right),
\label{ansatz}
\end{equation}
where $C$ is a contour in the complex $\omega$-plane.
This ansatz is motivated by the integral representation of the {Bessel
function} $J_0$ at $\al=0$ ($\tgr\Longrightarrow F=1$).

The substitution into Eq.~(\ref{edifladder2}) reduces it to the
hypergeometric equation ($\xi =\al S/T$)
\begin{equation}
\xi(1+\xi) F''_{\xi\xi}+ [1+\sqrt{\bl}(\omega+\omega^{-1})](1+\xi)F'_\xi
+\bl F=0
\label{hg}
\end{equation}
whose solution is given by {hypergeometric functions}.
How to draw the contour $C$ to satisfy the boundary conditions~\rf{bc} 
is described in Appendix~C.

A great simplification of the solution~\rf{ansatz} occurs at $S=T$,
$a=b$ and $\al=-1$:
\begin{equation}
\G_{\al=-1} (T,T ;a,a)=
\frac{1}{\sqrt{\bl \tau (\tau - 2 \pi i)}} \,
J_1 \left(2 \sqrt{\bl \tau (\tau - 2 \pi i)}   \right)
\label{GfinJ}
\end{equation}
with
\be
\ln \frac Ta=\tau\,, \qquad
\ln \left(-\frac Ta\right)= \tau - i\pi
\ee
and $\bl$ given by \eq{bl}.

In Appendix~B this Bessel function is reproduced in the double-logarithmic 
approximation. It is similar to that obtained by {Erickson, 
Semenoff, Zarembo (2000)} for a {circular Wilson loop}.
It is $J_1$ rather than $I_1$ because of Minkowski space.

Nothing good happens with the contribution of ladders 
to the cusp anomalous dimension.
It is {\it {not}}\/ of the form prescribed by the {renormalizability}
(cf.\ \eq{Km}):
\be
W\left( \tgr{\Gamma}_{\rm l.c.}\right) \propto \e^{-\frac 14 f(\bl) \ln ^2 
  \frac T{\tcy a}} \,.
\ee
A miniconclusion of this fact is that diagrams with interaction have 
to contribute.

\tit{Two-loop ladder diagram}

\begin{figure}[h]
\vspace*{3mm}
\centering{\includegraphics[width=5cm,height=5cm]{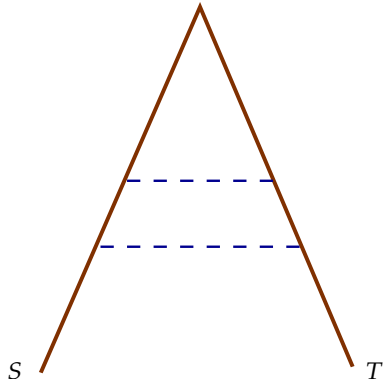}  
}
\caption[ladder diagram]  
{Ladder diagram with two rungs.}
\label{fi:ladd2}
\end{figure}

The contribution to the cusp anomalous dimension 
of the ladder diagram with two rungs depicted in Fig.~\ref{fi:ladd2}
was calculated by
{Korchemsky, Radyushkin (1987)}. 

The result is
\bea
\gamma_{\rm cusp}^{\tma{\rm (lad)}}&=&
\frac{\lambda^2}{128\pi^4} \frac{(\cosh\tre\theta-1)^2}{\sinh^2\tre\theta} 
\int_0^\infty \frac{d\sigma}{\sigma}
\ln\left( \frac{1+\sigma \e^{\tre\theta}}{1+\sigma \e^{-{\tre\theta}}}\right)
\ln\left( \frac{\sigma +\e^{\tre\theta}}{\sigma +\e^{-{\tre\theta}}}  \right)
\non&\ra&
\frac{\lambda^2}{96\pi^4} \left( 
\tre\theta^3 +\frac {\pi^2} 2 \tre\theta + {\cal O}(1) \right).
\label{lad2}
\eea 

The 
$\tre\theta^3$ -term {should be cancelled by interaction}!
Therefore, not only ladder diagrams are essential to order $\lambda^2$.

Similar results hold for the {light-cone} Wilson loop, when
\be
\G^{\rm ladd.}_{\rm l.c.}=1- \frac {\bl}{2}\ln^2\frac T{\tcy{\varepsilon}} 
+\frac {\bl^2}{12}\ln^4\frac T{\tcy{\varepsilon }}-
\frac {\bl^2 \pi^2}{12}\ln^2\frac T{\tcy{\varepsilon}} \,.
\ee
The term $\ln^4\frac T{\tcy{\varepsilon }}$ 
is again to be cancelled by diagrams with interaction.

\tit{Anomaly surface term}

\begin{figure}[h]
\centering{\includegraphics{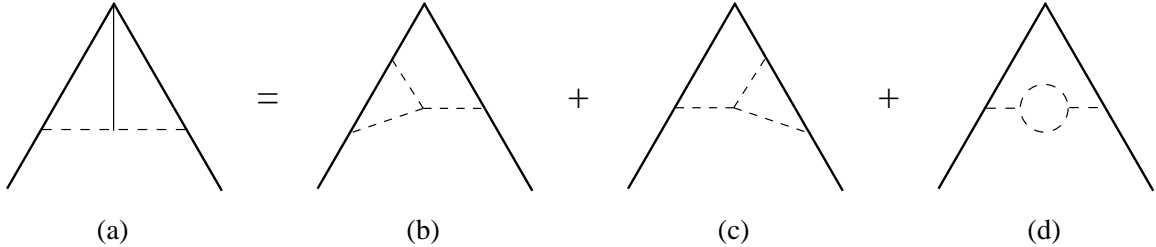}  }
\caption[ladder diagram]  
{Interaction diagrams of the order $\lambda^2$. The sum of the diagrams 
(b), (c) and (d) on
the right-hand side is equal to the anomalous term represented by
the diagram (a) on the left-hand side.}
\label{fi:anom2}
\end{figure}

As was shown by {Y.~M., Olesen, Semenoff (2006)}, the 
cancellation between the diagrams with the three-gluon vertex and  the
corrections to propagators (both are of the order $\lambda^2$) is 
{\em not}\/ {complete}. These diagrams are depicted in Fig.~\ref{fi:anom2}.

The cancellation of the diagrams 
(b), (c) and (d) on
the right-hand side is not complete as it is 
for a straight line or a circular
loop. A nonvanishing surface term comes from integration by parts.
It is represented by
the diagram (a) on the left-hand side.
It gives the following contribution to the cusp anomalous dimension
\bea
\gamma_{\rm cusp}^{\tma{\rm anom}}&=& -\frac{\lambda^2}{16\pi^4} 
\frac{\cosh \tre\theta -1}{\cosh \tre\theta} 
\Bigg(
\int_0^{\tre\theta} + \int _0^{\pi/2}\Bigg)
\frac{ \d\psi \, \psi}{1- \cosh^2 \psi/\cosh^2 \tre\theta} 
\ln \frac{\cosh^2 \tre\theta}{\cosh^2 \psi} 
\non &\ra & -\frac{\lambda^2}{96\pi^4} 
\left( \tre\theta^3 +\pi^2\tre\theta +{\cal O}(1) \right) .
\label{anom2}
\eea

The $\tre\theta^3$ -terms are mutually cancelled 
in the sum of the contributions of the ladder diagram~\rf{lad2}
and the anomaly diagram~\rf{anom2}. The remaining linear-in-$\tre\theta$ 
term {reproduces} the known results
\be
\gamma_{\rm cusp}= \frac {\tre\theta} 2
\left(\frac{\lambda}{2\pi^2} -\frac{{\lambda}^2}{96\pi^2}\right)
+{\cal O}(\theta^0)
\ee
for the {two-loop cusp anomalous dimension}.


Both the ladder contribution~\rf{lad2} and the anomaly contribution~\rf{anom2}
simplify at the light cone. 
To demonstrate the exponentiation to the order $\lambda^2$,
it is convenient to apply the so-called
``non-Abelian exponentiation theorem'' which states that
\be
\ln W = \G^{(1)}+A^{(2)}-\G^{(2)}_{\rm crossed} +{\cal O}(\lambda^3)\,,
\ee
where $\G^{(2)}_{\rm crossed}$ denotes the ladder diagram with two {\em crossed}\/
rungs, which is nonplanar and have to be added (and correspondingly 
subtracted) for the exponentiation of the diagram 
in Fig.~\ref{fi:1-loop}(b) of the order $\lambda$. 

We calculate the difference 
 of the anomaly and crossed
ladder diagram using the regularization via dimensional reduction to
$d=4-\epsilon$:
\be
A^{(2)}-\G^{(2)}_{\rm crossed}=2\bl^2\int_a^S\frac{\d s}{s^{1-\epsilon}} 
\int_b^T \frac{\d t}{t^{1-\epsilon}} 
\left[\frac{\pi^2}6- {\rm Li_2}\Big(\frac{\alpha s}{\alpha s+t} \Big) \right].
\label{A-G}
\ee

The integral in \eq{A-G} is fast convergent when $s\to\infty$
or $t\to0$ because $
 {\rm Li_2}(1)={\pi^2}/6$.
Alternatively, the second term (dilogarithm) can be omitted 
with the double-logarithmic accuracy for
$\alpha s\ll t$. This justifies the exponentiation to the order $\lambda^2$ and
 reproduces the two-loop anomalous dimension because only the domain
$\alpha s\la t$ is essential both for $\alpha S\ll T$ and $\alpha S\gg T$.

\tit{Higher-order anomaly terms}

A question arises as to whether the anomaly
surface term of order $\bl^2$ is the only one
(like an anomaly in QFT) or next order anomaly terms also appear.
This question can be answered in the double logarithmic approximation.

\begin{figure}[h]
\centering{\includegraphics[width=5cm,height=5cm]{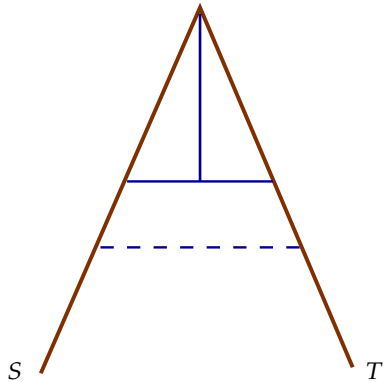}  }
\caption[anomaly diagram]  
{Dressing of the anomaly diagram of the order $\lambda^2$ by a ladder.}
\label{fi:dress2}
\end{figure}

Let us consider the sum of the ladder diagram with three rungs and
the anomaly diagram of the order $\lambda^2$ dressed by a ladder as
is depicted in Fig.~\ref{fi:dress2}.
It is easy to see that this sum does not provide
the coefficient required for the exponentiation to the order $\lambda^3$,
like in 
\be
W_{\rm l.c.}\left( \tgr{\Gamma} \right) = \e^{-\frac \bl2 \Tau^2}\,, \qquad
\al S \ga T \,,
\label{abel}
\ee
which itself is a consequence of a dual conformal symmetry by
Drummond, Korchemsky, Sokatchev (2008),
Drummond, Henn, Korchemsky, Sokatchev (2008) and reviewed in
the lectures by G.~Korchemsky at this School. 

\begin{figure}[h]
\centering{\includegraphics[width=5cm,height=4.5cm]{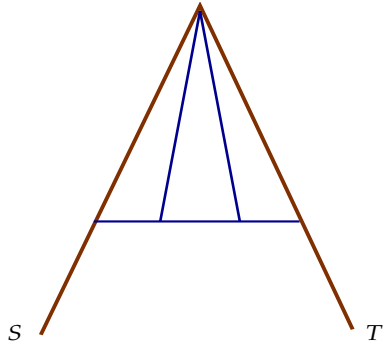}  }
\caption[anomaly diagram]  
{New anomaly diagram of the order $\lambda^3$ which provides the exponentiation
in the double logarithmic approximation.}
\label{fi:nextanom}
\end{figure}

However, to guarantee the exponentiation
in the double logarithmic approximation to the order $\lambda^3$,
it is enough to add a new anomaly
diagram depicted in Fig.~\ref{fi:nextanom}. It appears from a particular
class of interaction diagrams of the order $\lambda^3$ after two integrations
by parts.

\tit{Higher-order anomaly terms (continued)}

\begin{figure}[h]
\vspace*{3mm}
\centering{
{\includegraphics[width=6cm]{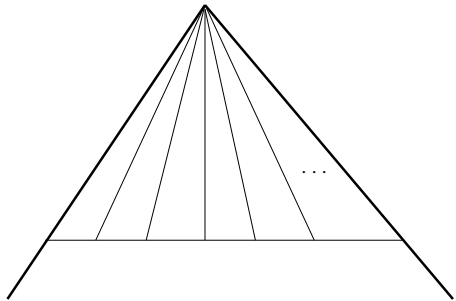}} 
}
\caption[Anomaly diagrams]
{Particular class of anomaly diagrams.}
   \label{fi:pie}
\end{figure}

The exponentiation to the order $\lambda^3$, described in the previous Section,
prompts to analyze 
 anomaly diagrams of 
the type depicted in Fig.~\ref{fi:pie}, \ie of the type of a sliced pie.

This class of diagrams can be exactly calculated recursively.
Let us denote $x=-u s$ and $y=v t$ with $y^2=0$ and $x^2\neq 0$ 
to proceed recursively.
We find for the $n$-loop ($n$-slice) diagram
\bea
P^{(n)}(x,y)&=&  2^{n-1}(-1)^n \frac{\bl^n}{4(n-1)!}  
\left(\frac4\epsilon\right)^{n-1}
\frac{\Gamma(1-\frac{n\epsilon}2)}
{\Gamma^n(1-\frac\epsilon2)}\non &&~~~
\times u\cdot v
\int_0^S \d s \int_0^T \d t \int_0^1 \d\tau_1\cdots\d\tau_{n-1}\, 
\frac{\prod_{k=1}^{n-1} \tau_k^{-k\epsilon/2}(1-\tau_k)^{-\epsilon/2}}
{\left( x^2-2 \prod_{k=1}^{n-1} \tau_k \,x\cdot y \right)^{1-n\epsilon/2}}\,.
\non &&
\label{er}
\eea
We have inserted a combinatorial factor of $2^{n-1}$ because the $n-1$ lines,
that are trapped by the cusp, may come from both sides.

Equation~\rf{er} is the exact result for the loop with 1 light-cone segment.

For the loop with 2 light-cone segments, we put $x^2=0$ after which
the integral is expressed via $\Gamma$-functions:
\bea
P^{(n)}&=& (-1)^n \frac{\bl^n}{16(n!)^2}  
\left(\frac4\epsilon\right)^{2n}
\frac{\Gamma(1-\frac{n\epsilon}2)\Gamma(1+\frac{(n-1)\epsilon}2)}
{\Gamma(1-\frac\epsilon2)}\left(2u\cdot v ST\right)^{n\epsilon/2}\,. 
\eea
It gives again the Bessel function $J_0$ in the double logarithmic
approximation rather than the exponential~\rf{abel}.

In fact this illustrates why it is very difficult to obtain
the exponential~\rf{abel} in the framework of planar QFT,
where the appearance of Bessel functions is instead quite natural and
understandable from the relationship between connected planar
and all planar diagrams. The exponential~\rf{abel} is rather
quite natural for Abelian theories when planar and nonplanar
diagrams are both essential. Actually the role of the anomaly
terms is simply to complete the Bessel function to an exponential.

A question immediately arises as to what 
is the equation which sums planar diagrams: ladders and anomalous 
to provide the exponentiation in the double logarithmic approximation? 
For this purpose we shall consider in the next lecture
the cusped loop equation.

\newsection*{Lecture 3.~~Cusped Loop Equation}
\setcounter{section}{3}
\setcounter{subsection}{0}

In this lecture I consider the loop equation for cusped Wilson loops.
I begin with a review of the modern formulation of the loop equation,
describe its supersymmetric extension and a UV regularization.
Then I concentrate on
specific features of the loop equation for
cusped loops and show how to extract the cusp anomalous dimension from the 
loop equation.

\tit{Loop equation in QCD}

The {Schwinger--Dyson equation} of Yang--Mills theory
\be
\nabla_\mu^{ab} F_{\mu \nu }^b( x)  
\stackrel{\rm w.s.}{=}  \hbar \,\frac \delta {\delta
A_\nu^a ( x) } \,, 
\label{S-D}
\ee
when applied for Wilson loops, was translated by Y.~M., Migdal (1979)
to the {\em loop equation}\/ which is a closed equation as $\N\to\infty$:
\be
\partial _\mu ^x\frac \delta {\delta \sigma _{\mu \nu }( x)
}W( \tgr{C})  =  \lambda\oint\limits_{\tgr{C}} \d y_\nu \,
\delta^{(d)}\! \left( x-y\right)
W( \tgr{C}_{yx})\, W( \tgr{C}_{xy})\,.
\label{LE}
\ee
This original loop equation
 includes the operators of {\em path}\/ and {\em area}\/ derivatives,
which are defined for functionals of Stokes type obeying
the zig-zag symmetry.
The product of two $W$'s on the right-hand side is due to the 
large-$\N$ factorization. 

The following vocabulary for translation from the ordinary space to loop space
is in order.

\mbox{}
\hspace*{2cm}\mbox{}

\centerline{
\begin{tabular}{|rl|rl|} \hline 
\multicolumn{2}{|c|} {\tma{Ordinary space}}   & 
\multicolumn{2}{c|} {\tma{Loop space}}  \\ \hline \hline  
\mbox{}& & & \\
$\Phi[ A ]$ & \tbr{Phase factor} &  $\Phi( \tgr{C} )$  
 & \tbr{Loop functional} \\ 
\vspace*{-2mm} 
\mbox{} & & & \\
$F_{\mu\nu}(x)$& \tgo{Field strength} & 
$\displaystyle \frac{\delta}{\delta \sigma_{\mu\nu}(x)}$ & 
 \tgo{Area derivative}\index{area derivative} \\ 
\vspace*{-2mm}\mbox{} & & & \\ 
$\nabla_\mu^x$& \tgr{Covariant derivative}& 
 $\partial_\mu^x$ &  \tgr{Path derivative}\index{path derivative} \\ 
\mbox{} & & & \\
$\nabla \wedge F =0$& \tcy{Bianchi identity}\index{Bianchi identity} & 
  &  \tcy{Stokes functionals}\index{Stokes functional}  \\ 
\mbox{} & & & \\
$-\nabla_\mu F_{\mu\nu}$\hspace*{.5cm} &\tbl{Schwinger--Dyson} & 
  &  \tbl{Loop}\index{Schwinger--Dyson equation!Yang--Mills theory}   \\ 
$=\delta/\delta A_\nu$&  \tbl{equation} & 
  &   \tbl{equation} \\ \mbox{} & & & \\\hline 
\end{tabular} }
\vspace*{5mm}

\tit{Loop-space Laplace equation}

A very nice (and equivalent!) form of the loop equation can be 
obtained by  one more contour integration over $x$: 
\be
\Delta W(\tgr{C}) 
= \tgo{\lambda} \oint\limits_{\tgr{C}}\d x_\mu \oint\limits_{\tgr{C}}\d y_\mu\, 
\delta^{(d)}(x-y) \, W(\tgr{C}_{yx})\, W(\tgr{C}_{xy}) \,.
\label{LLE}
\ee

The operator $\Delta$ on the left-hand side of \eq{LLE} is nothing
but the loop-space Laplacian
\be
\Delta \equiv
\oint\limits_{\tgr{C}} 
\d x_\nu \,\partial_\mu^x \,\frac{\delta}{\delta\s_{\mu\nu}( x)}
 =\int\limits_{\s_i}^{\s_f} \d\s  \int\limits_{\s-0}^{\s+0} \d\s^\p 
\frac{\delta}{\delta x_\mu(\s^\p)} \frac{\delta}{\delta x_\mu(\s)}
\label{Lapl}
\ee
which is a proper functional extension of a finite-dimensional
Laplacian that respects continuity of the loop. 
As is seen from the right-hand side of \eq{Lapl}, the loop-space Laplacian 
is defined for a {much wider} class of functionals than Stokes functionals.
This will be important for a supersymmetric extension of the loop equation.

The loop-space Laplace equation~\rf{LLE} 
is associated with the {second-order} Schwinger--Dyson equation
\be
\int \d^dx\,\nabla _\mu F_{\mu \nu }^a\left( x\right) \frac \delta {\delta
A_\nu ^a( x) }  
\stackrel{\rm \tre{w.s.}}{=}  
\hbar \int \d^dx\,\d^dy\,\delta^{(d)}\! 
\left(
x-y\right) \frac \delta {\delta A_\nu ^a( y) }\frac \delta
{\delta A_\nu ^a( x) } 
\ee
in the same sense as the original loop equation~\rf{LE} is
associated with \eq{S-D}.
This fact was utilized by Halpern, Y.~M. (1989) to construct a
 non-perturbative gauge-invariant regularization of \eq{Lapl} by  
substituting
\be
\delta^{bc}\delta^{(d)}(x-y) \stackrel{\rm \tre{reg.}}{\Longrightarrow}
\LA y\left|
\left( \e^{{\tcy{a}^2 \nabla^2}/{2}}  \right) ^{bc} \right| x \RA \,,
\ee
where $\tcy{a}$ is a UV cutoff.

\tit{Smearing of loop-space Laplacian}

A smearing of the loop-space Laplacian 
is needed to {invert} it, \ie
 to produce a {Green function}. 

The proper smearing procedure (which makes a second-order operator 
from the first order loop-space Laplacian) reads as
\begin{eqnarray}
\lefteqn{\Delta^{(\tcy G)} =
\int\limits_0^1\d\sigma \int\limits_0^1 \d\sigma ^\prime\,  
G(\sigma ,\sigma ^\p)
 {\delta \over \delta x_{\mu }(\sigma ^\prime )} {\delta \over \delta x_{\mu 
}(\sigma )} } \non 
&=&\int\limits_0^1 \d\sigma \pintab \d\sigma^\prime\, G(\sigma ,\sigma ^\p)
 {\delta \over \delta x_{\mu }(\sigma ^\prime )} {\delta \over \delta x_{\mu 
}(\sigma )} + \Delta 
\end{eqnarray}
with the {parametric-invariant} smearing function
\be
G(\s_1,\s_2) = \e^{-{\big|\int_{\s_1}^{\s_2}}
\d \s \sqrt{\dot x^2(\s)}\big|/{\tcy\varepsilon}} 
\quad \quad(\tcy\varepsilon \ll L={\rm length})\,.
\ee
Here $\tcy\varepsilon$ has the meaning of a {\em stiffness}\/ of the loop.

\tit{Green function of functional Laplacian}

Loop-space Laplacian was {inverted} by Y.~M. (1988)
 to produce a {Green function} which is 
useful for an iterative solution.

The functional Laplace equation 
\be
\Delta ^{(\tcy G)} W [x] = J [x]
\label{fLE}
\ee
with the proper choice of boundary conditions can be solved 
for a given $J[x]$ to give
\be
W [x] = 1 - {1\over 2}\int_0^\infty \d A \left\{\LA J [x+\sqrt{A}\tgr\xi ] 
\RA_{\tgr\xi}^{(\tcy G)}
- \LA J [\sqrt{A}\tgr\xi ] \RA_{\tgr\xi }^{(\tcy G)} \right\}\,.
\label{GF}
\ee

The average over the loops $\tgr\xi(\s)$ in \eq{GF}
is given by the path integral
\be
\left\langle F [\tgr\xi] \right\rangle_{\tgr\xi}^{(\tcy G)} =
 \frac{\int_{\tgr\xi(0)=\tgr\xi(1)} D\tgr\xi \e^{-S} F [\tgr\xi]}
{\int_{\tgr\xi(0)=\tgr\xi(1)} D\tgr\xi \e^{-S}}
\label{PI}
\ee
over closed trajectories with the local action  
\be 
S = {1\over 4}\int_0^1 \d\sigma  \Big\{ 
\frac{\tcy\varepsilon}{\sqrt{\dot x^2(\s)}}\, \dot{\tgr\xi}^2 (\sigma )  +
\frac{\sqrt{\dot x^2(\s)}}{\tcy\varepsilon}\, \tgr\xi^2 (\sigma )\Big\} \,.
\label{Sa}
\ee
This extends the results of the French 
mathematician R.~G\^ateaux (early 1900's) 
for the functional Laplacian (obtained for the case of $L_2$ space 
of functions with an integrable square)
to the case of loop space when loops
are always continuous functions. It is immediately seen from the action~\rf{Sa}
that the presence of the stiffness $\varepsilon$ is crucial to have
a Wiener-type measure in the path integral~\rf{PI} and correspondingly
continuous trajectories.

\tit{Iterative solution}

In large-$\N$ Yang--Mills theory the {regularized}
$J[x]$ is as above {bilinear} in $W$:
\bea
J^{(\tcy G)}[x] &= & \lambda \int_0^1\int_0^1 \d\s_1 \d\s_2 \,(1-G(\s_1-\s_2))\,
\dot{x}^\mu(\s_1) \dot{x}^\mu(\s_2) \non &&~\times
 \int_{r(0)=x(\s_1)}^{r(\tcy\ca^{2})=x(\s_2)} {\cal D} r
\e^{-\frac 12 \int_0^{\ca^{2}} \d \tau \,\dot{r}^2(\tau)}\non &&~~~~~~ \times
W(\tgr C_{x(\s_1)x(\s_2)}r_{x(\s_2)x(\s_1)}) \,
W(\tgr C_{x(\s_2)x(\s_1)}r_{x(\s_1)x(\s_2)})\,.
\label{Jreg}
\eea

It can been shown that an 
iterative solution in $\tgo\lambda$ starting from $W_{0}(C)=1$
recovers Yang--Mills perturbation theory.
All that can be deduced from the general formula 
\be
\LA\e^{\i\sqrt{A} \int \d\sigma \dot p(\sigma )
\tgr\xi (\sigma )}\RA_{\tgr\xi}^{(\tcy G)} 
= \e^{-{A} \int \d\sigma \int \d\sigma ^\prime \dot p(\sigma )G(\sigma
-\sigma ^\prime )\dot p(\sigma ^\prime )/2} \,,
\ee
where $p^{\mu }(\sigma )$ ($p^{\mu }(0)=p^{\mu }(1)$)  represents  a 
{momentum-space loop}.
In particular, 
the {triple gluon} vertex remarkably appears from doing an uncertainty 
$\tcy \varepsilon \times 1/\tcy \varepsilon$.

\tit{SYM loop equation}

An extension of the loop equation~\rf{LLE} to
${\cal N}=4$ SYM was proposed by {Drukker, Gross, Ooguri (1999)}.
An equivalent equation was derived earlier by 
Fukuma, Kawai, Kitazawa, Tsuchiya (1998) in connection
with the IIB matrix model.

The equation closes for general supersymmetric loops
\be{\bbox C}=\{x_\mu(\sigma), Y_i(\sigma);
\zeta(\sigma)\}   \qquad \mu=1,\ldots,4,~i=1,\ldots,6\,,
\label{boldC}
\ee 
where $\zeta(\sigma)$ denotes the Grassmann odd component.
An ${\cal N}=4$ supersymmetric extension of the loop-space Laplacian~\rf{Lapl} 
is
\be 
{\bbox \Delta} = \lim_{\eta\to0} \int \d s
\int_{s-\eta}^{s+\eta} \d s' \left( \frac{\delta^2}{\delta
x^\mu(s')\delta x_\mu(s)}+ \frac{\delta^2}{\delta Y^i(s')\delta
Y_i(s)}+ \frac{\delta^2}{\delta \zeta(s')\delta \bar\zeta(s)}\right).
\ee
To return to the SYM Wilson loops~\rf{WSYM}, we have to put
 $\dot Y^2=\dot x^2$, $\zeta=0$ after acting by
${\bbox \Delta} $. We use the bold ${\bbox C}$ for general loops~\rf{boldC}
and normal $\tgr{C}$ for SYM loops~\rf{9-dim}. 

The resulting  ${\cal N}=4$ SYM loop equation then reads 
\bea
\bbox{\Delta} \ln W(\bbox{C})\Big|_{\bbox{C}=\tgr{C}}&= &
\lambda \int \d \sigma_1 \int \d \sigma_2 \,
\left( \dot x_\mu(\sigma_1)\dot x_\mu(\sigma_2)-
|\dot x_\mu(\sigma_1)||\dot x_\mu(\sigma_2)|
\right) \non &&~~~~~\times
\delta^{(4)}{(x_1-x_2)}
\frac{W(\tgr{C}_{x_1 x_2})W(\tgr{C}_{x_2x_1})}{W(\tgr{C})}\,.
\label{SYMLE}
\eea
For latter convenience we have applied ${\bbox \Delta}$ to $\ln W(\bbox{C})$
and used the fact that ${\bbox \Delta}$ is a first-order operator (obeys 
the Leibnitz rule). The right-hand side is correspondingly
divided by $W(\tgr{C})$.

Probably a simpler version of ${\cal N}=4$ SYM loop equation exists
which is written directly for SYM loops $W(\tgr{C})$ along the line of
the derivation of the loop equation for an Abelian scalar loop by
Y.~M. (1988). It is crucial for this approach that the loop-space Laplacian
is well-defined for the functionals of the type~\rf{WSYM} which do {\em not}\/
obey the zig-zag symmetry because of the presence of $\sqrt{\dot x^2}$.
But \eq{SYMLE} will be enough for the purposes below. 

\tit{Cusped loop equation}

The loop equation for cusped Wilson loops, which we denote by $\tgr\Gamma$,
can be obtained by substituting $\tgr C=\tgr\Gamma$ in \eq{SYMLE}.
We then obtain the following cusped loop equation
\bea
\bbox{\Delta} \ln W(\bbox{C})\Big|_{\bbox{C}=\tgr\Gamma}&= &
\lambda \int \d \sigma_1 \int \d \sigma_2 \,
\left( \dot x_\mu(\sigma_1)\dot x_\mu(\sigma_2)-
|\dot x_\mu(\sigma_1)||\dot x_\mu(\sigma_2)|
\right) \non &&~~~~~\times
\delta^{(4)}{(x_1-x_2)}
\frac{W(\tgr\Gamma_{x_1 x_2})W(\tgr\Gamma_{x_2x_1})}{W(\tgr\Gamma)}\,.
\label{cuspedLE}
\eea

The cusped loop equation~\rf{cuspedLE} has several remarkable properties.
One of them is the following. The right-hand side of the usual Yang--Mills loop
equation~\rf{LLE} is of the order $L/\tcy a^3$, where $\tcy a$ is a UV cutoff.
For the  ${\cal N}=4$ SYM loop equation~\rf{SYMLE} this term is cancelled
by scalars so the right-hand side of \eq{SYMLE} is
 $\sim (L\tcy a)^{-1}$ for \tma{smooth} loops.  
But it is easy to estimate that it is 
$\sim \tcy a^{-2}$  for {cusped} loops which is larger. Therefore,
some interesting information about cusped Wilson loop can be
extracted from the loop equation at the order $\sim \tcy a^{-2}$.
As we shall now see, this is the cusp anomalous dimensions.

A crucial observation is the following relation that holds for 
cusped Wilson loops:
\be
\Delta \ln W(C)\Big|_{C=\tgr\Gamma} = -
\frac{\d}{\d \tcy a^2} \ln W(\tgr\Gamma) + {\cal O}(\tcy a^{-1})\,. 
\label{Dda}
\ee
Here the differential operator on the right-hand side which
 is, in general, regularization-dependent is written
for the Schwinger proper time regularization, like in \eq{Jreg}. 
The relation~\rf{Dda} can be verified in
perturbation theory for usual Yang--Mills and most probably extends
to the SYM case.

Noting that the term of the order $\tcy a^{-2}$ on the right-hand side
of \eq{Dda} gives the cusped anomalous dimension owing to \eq{defgamma},
we find that the cusp anomalous dimension equals to the term
of the order $\tcy a^{-2}$ on the right-hand side of the 
(regularized) loop equation:
\bea
{
\frac{2}{\tcy a^2} \gamma_{\rm cusp}\left(\theta,\lambda\right)}
+{\cal O}\left(\tcy a^{-1}\right)
&= &\lambda
\int \d \sigma_1 \int \d \sigma_2 \,
\left( \dot x_\mu(\sigma_1)\dot x_\mu(\sigma_2)-
|\dot x_\mu(\sigma_1)||\dot x_\mu(\sigma_2)|
\right) \non &&~~~~~\times
\delta^{(4)}_{\ca}(x_1-x_2)
\frac{W(\tgr\Gamma_{x_1 x_2})W(\tgr\Gamma_{x_2x_1})}{W(\tgr\Gamma)}
 \,.
\label{gammaE}
\eea
Here the right-hand side is to be regularized according to \eq{Jreg}.

This very interesting fact was first observed to the order $\lambda$ of
perturbation theory by {{Drukker, Gross, Ooguri (1999)}} and
{verified} to order $\lambda^2$ for arbitrary $\tre\theta$ by
{{Y.~M., Olesen, Semenoff (2006)}}.
The nontrivial function of $\tre\theta$ thus reproduced to the order $\lambda^2$
is the sum of
the contribution of the ladder diagram~\rf{lad2} and the anomaly 
diagram~\rf{anom2}.

The contribution of the {ladder diagram} of order $\lambda^2$ straightforwardly 
comes iteratively by substituting the ladder
diagram of order $\lambda$ into the right-hand side of \eq{gammaE}.
A planar part is cancelled between the numerator and the denominator on
the right-hand side of \eq{gammaE} so what is left is analogous to
the diagram with two crossed ladders (with the minus sign) as in \eq{lad2}.
This is a well-know implementation of the ``non-Abelian exponentiation 
theorem'' in the loop equation. 

Alternatively, the contribution of
the anomaly diagram is reproduced in a very nontrivial way,
when a gluon is attached to the 
{regularizing path} $r_{x_1 x_2}$. While the length of this path is
$\sim \tcy a$, \ie very small, this smallness is compensated by
a very large factor. The calculation is performed using an  
important formula of the loop dynamics derived by
{{Y.~M., Migdal (1981)}}:
\bea
\lefteqn{\hspace*{-1cm}\int\limits_{z(0)=x \atop z(\tau)=y}
{\cal D}z(t)\e^{-\int_0^\tau \d t \,\dot z^2(t)/2}
{\int^y_x \d z^\mu \delta^{(d)}(z-u)}= 
\int_0^\infty \d \tau_1 \int_0^\infty \d \tau_2\,
\delta \left(\tau -\tau_1-\tau_2\right)} \non
&&\hspace*{3cm}\times \frac{1}{(2\pi \tau_1)^{d/2}}\e^{-(x-u)^2 /2 \tau_1}
\frac{\stackrel{\leftrightarrow}{\partial} }{\partial u_\mu}
\frac{1}{(2\pi \tau_2)^{d/2}}\e^{-(y-u)^2 /2 \tau_2} \,.
\eea
This formula in the usual
Yang--Mills theory was used to reproduce the three gluon vertex.
Now in ${\cal N}=4$ SYM it reproduces directly the anomaly diagram,
rather than individual interaction diagrams.

An expectation is that {the loop equation may be useful for 
analyzing next orders in $\lambda$} and, in particular, to verify
the exponentiated solution~\rf{abel} in the double logarithmic approximation.

\tit{Some comments about large-$\N$ QCD}

Some of the results described above for cusped Wilson loops 
in ${\cal N}=4$ SYM are applicable also to large-$\N$ QCD.
Actually, nobody considered before specific features of the loop
equation for cusped loops.

A first immediate consequence of the cusped loop equations is
that $|\dot x|$ on the right-hand side (which comes from scalars)
{can be neglected near the light-cone}, reproducing 
the {same cusped loop equation as in QCD}.

{This may indicate that} $\gamma_{\rm cusp}$ coincide in both cases
while  the
difference is absorbed  by the charge renormalization which is 
present in QCD and missing in ${\cal N}=4$ SYM.
{This may be understood because supersymmetry is broken by construction} 
in {the presence of a cusp} and the larger the cusp the larger is the breaking.
However, this assertion is in fact rather vague because the cusp anomalous
dimension is regularization-dependent. But this indeed works to the
order $\lambda^2$, if the regularization in both cases is via dimensional 
reduction. A comparison of recent explicit calculations of the anomalous 
dimensions
of twist-two operators to the order $\lambda^3$ in QCD by
Moch, Vermaseren, Vogt (2004) with those in SYM by 
Kotikov, Lipatov, Onishchenko, Velizhanin (2004)
may be useful for this purpose.

\section*{Conclusions}

I list below a (quite incomplete) set of conclusions from these lectures:

\begin{itemize}
\addtolength{\itemsep}{-5pt}
\item {Cusped Wilson loops are convenient for
studies of the anomalous dimensions;}

\item {The minimal surface of an open string in $AdS_5\otimes S^5$
determines the cusp anomalous dimension
at large $\lambda$ via the AdS/CFT correspondence;}

\item {Ladder diagrams themselves do not give a reasonable result for
the cusp anomalous dimension, so that diagrams with interaction are
essential;}

\item {A cancellation of interaction diagram to the order $\lambda^2$ 
is not complete for 
${\cal N}=4$ SYM cusped Wilson loops and an anomaly surface term remains;}

\item {Results in the double logarithmic approximation 
indicate that higher-order interaction diagrams are
 also essential for the exponentiation;}

\item {The loop equation has specific features for cusped loops when its
right-hand side reproduces the cusp anomalous dimension;}

\item {There are indications that some of the results about the cusp anomalous 
dimension in ${\cal N}=4$ SYM could persist for QCD;}

\item {A challenging problem is to obtain $\sqrt{\lambda} $ for large  
$\lambda$ in perturbation theory.}
\end{itemize}

\eop

\setcounter{section}{0}
\setcounter{subsection}{0}
\setcounter{equation}{0}
\appendix{Minimal surface in $AdS_5\otimes S^5$}

I briefly review in this Appendix the calculation of the cusp
anomalous dimension at large $\lambda$ which is based on the 
open-string/Wilson-loop correspondence described in Sect.~\ref{1.10}.
Only the leading order calculation by 
Kruczenski (2002), Y.~M. (2003)
in the supergravity approximation
is considered.

\tit{Near cusp ansatz}

We describe the 
$AdS_5$ space by the Poincar\'e {coordinates}
\be
x^0=t\,\tre , \quad x^1=x\,\tre ,\quad x^2=x^3=0\,\tre , \quad z
\ee
choosing the $\Pi$-shaped loop to be located in the $0,1$-plane. 
The associated $AdS_3$ {metric} is
\be
\d s^2=\tgo {R^2} \frac{\d t^2-\d x^2-\d z^2}{z^2} \,.
\ee

We parametrize the string worldsheet by the coordinates
{$\tau=t$}, {$\sigma=x$} so the open-string action reads as
\be
A=- \frac{\tgo {R^2}}{2\pi \tgo {\alpha^\prime}}
\int\limits_0^y \d x \int\limits_x^\infty \d t \frac{1}{z^2}
\sqrt{1+\left(\frac{\partial z}{\partial x}\right)^2 -
\left(\frac{\partial z}{\partial t}\right)^2}\,.
\label{Aopen}
\ee
{It is to be minimized for the function} $z(t,x)$.

\begin{figure}[h]
\centerline{\includegraphics[width=10cm]{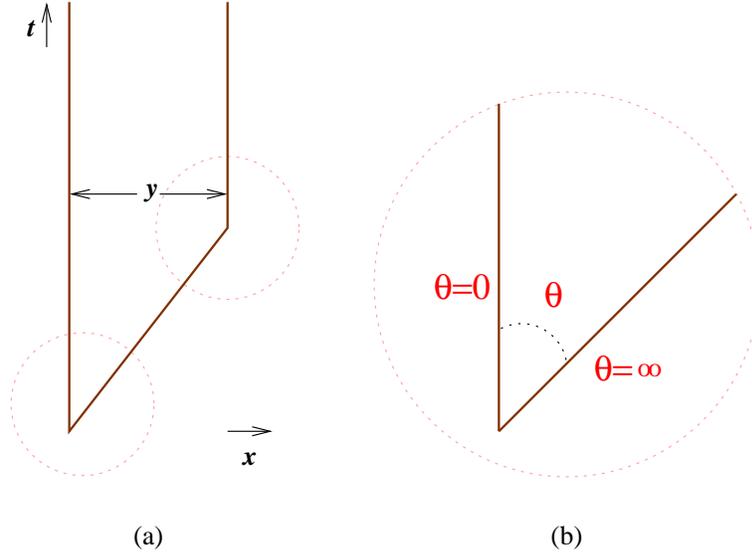}} 
\caption[one-loop perturbation theory]  
{Near cusp domains of the minimal surface
which determines the cusp anomalous
dimension. In Fig.~(b) one of the two domains is magnified.}
\label{fi:near-cusp}
\end{figure}

Using scale and Lorentz invariance (at the worldsheet) the following
ansatz was proposed by {Drukker, Gross, Ooguri (1999)} near the cusp 
to simplify the problem of minimizing the action~\rf{Aopen}:
\be
z(t,x)= \sqrt{t^2-x^2} \frac{1}{f(\tre \theta)}\,,\qquad 
\tre \theta = {\rm arctanh}\; \frac x t \,.
\ee
It works for the domain near the cusp depicted in Fig.~\ref{fi:near-cusp}.
As we shall see shortly it is the domain which contributes
to the cusp anomalous dimension at large $\tre \theta$ and
at the light cone.

The original two-dimensional variational problem is thus reduced
to a one-dimensional Euler--Lagrange one.

\tit{Euler--Lagrange problem}

We are led to minimize the one-dimensional action
\be
A = - \tgr2\frac{\tgo {R^2}}
{2\pi \tgo {\alpha^\prime}}\int \frac {\d x}{x} 
\int \d \tre\theta\, \sqrt{f^4-f^2 +\left. f ^\prime \right.^2} \,,
\label{E-L}
\ee
where {the factor of} $\tgr2$ {is because the $\Pi$-shaped
Wilson loop in Fig.~\ref{fi:near-cusp}(a) has two cusps}.

The standard condition that $\tre{z=0}$ {in the boundary}
is transformed into the boundary condition for $f(\theta)$: {$f(0)=\infty$} 
and an arbitrary  value of $f(\infty)$ 
{because} $\sqrt{t^2-x^2}\rightarrow 0$ {as} 
{$\theta\rightarrow\infty$}. Also,
$f=0$ is the horizon and  $f=1$ is  the $AdS$ light cone given by
{$z^2=t^2-x^2$}.

To minimize the action~\rf{E-L}, we solve the following
 {Euler--Lagrange equation}
\be
f^{\prime\prime} V = f'^2 V' +\frac 12 V' V \,,\qquad \tgr{V=f^4-f^2}
\label{ELe}
\ee
with the described boundary conditions.

\tit{Solution for $f(\theta)$}

The equation~\rf{ELe} can be easily integrated 
utilizing the {conserved ``energy''}
\be
E=\frac{V}{\sqrt{f'^2+V}} \qquad \tre\Longrightarrow \qquad 
\tma{f'=-\sqrt{\frac{V^2}{E^2}-V}} \,.
\label{energy}
\ee
It is seen from these formulas that
$\tre{\theta_{\rm max}}$ is finite unless {$E^2=-1/4<0$} 
{(an instanton-like solution)}.

An {exact analytic solution} to \eq{energy} is
\be
\tre\theta= \sqrt{2} \,{\rm arctanh}\, \sqrt{2(1-f^2)} 
-{\rm arctanh}\,\sqrt{1-f^2}  -
\frac {\i \pi}{2} \left(\sqrt2 -1 \right) .
\label{sol}
\ee
It {obeys the boundary condition} \tre{$\theta=0$} at $f=\infty$
{and} $ \tre{\theta\ra\infty}$ approaching the light cone {as}
$f\ra f_{\rm min}=1/{\sqrt{2}} $.
 
Only a part of the surface near the light cone described by the equation
\be
z=\frac{\sqrt{t^2-x^2}}{f_{\rm min}}=\sqrt{2(t^2-x^2)}
\ee
will be essential for the cusp anomalous dimension.
It is the same (space-like) minimal surface as found by Kruczenski (2002)
for space-like Wilson loops, {when it was always real}.

\tit{Cusp anomalous dimension}

Substituting the solution~\rf{sol} into the action~\rf{E-L},
we find that the essential domain of integration  is  when 
\be 
f\ra f_{\rm min}=\frac{1}{\sqrt{2}}+\sqrt{2}\e^{-\sqrt{2}\theta_{\rm max}}\,,
\qquad\quad \theta_{\rm max}=\ln
\frac{2\sqrt{2}x}{\ca}
\ee
and $\tre{\theta \ra \theta_{\rm max}}$.
We get for the divergent part of the minimal area
\bea
A_{\rm min}&=& \frac{\tgo {R^2}}
{\pi \tgo {\alpha^\prime}}\int_\ca^y \frac {\d x}{x} 
\int_0^{\tre{\theta_{\rm max}}} \d \tre\theta\, 
\sqrt{f_{\rm min}^4-f_{\rm min}^2} \non 
&=& i\frac{\tgo {R^2}}
{\pi \tgo {\alpha^\prime}} \int_{\ca}^y \frac{\d x}x
\frac{\tre{\theta_{\rm max}}}2 
=i\frac{\tgo {R^2}}
{4\pi \tgo {\alpha^\prime}}\ln^2 \frac y\ca \,.
\eea

From the open-string/Wilson-loop correspondence, described in Sect.~\ref{1.10},
we obtain
\be
W_{\rm SYM}(\tgr{\Pi})=\e^{2 \i A_{\rm min}} \qquad
\hbox{with}\quad \sqrt{\lambda}=\frac{\tgo {R^2}}
{\tgo {\alpha^\prime}}.
\ee
Comparing with \eq{Km} we find 
\be
 f(\tgo{\lambda})= 
\frac{\sqrt{\lambda}}\pi \qquad {\rm large}~\lambda
\label{sqrtl}
\ee
which reproduces the result by Gubser, Klebanov, Polyakov (2002)
quoted in Sect.~\ref{GKP}.
It is obtained for open string in the supergravity {approximation}
while the original calculation dealt with closed string in the plane wave limit.

The asymptotic behavior~\rf{sqrtl} of the cusp anomalous dimension at
large $\lambda$ has been remarkably reproduced from the spin-chain equation 
of Beisert, Eden, Staudacher (2007) first numerically by
Benna, Benvenuti, Klebanov, Scardicchio (2007) and then analytically by
Kotikov, Lipatov (2007),
Basso, Korchemsky, Kotanski (2008). The latter authors constructed a systematic 
expansion of the cusp anomalous dimension in $1/\sqrt{\lambda}$, 
reproducing also the ${\cal O}(1)$ correction
by Frolov, Tseytlin (2002) obtained from closed stings.
The next order in $1/\sqrt{\lambda}$ also agrees with the recent 
superstring calculations by Roiban and Tseytlin (2007), (2008). 

\eop
\setcounter{subsection}{0}
\setcounter{equation}{0}
\appendix{Double-Logarithmic Accuracy}

I describe in this appendix how the double logarithmic accuracy 
works for lightcone Wilson loops to the leading order,
confirming in particular previously obtained results.

\tit{Double logarithms at the light cone}

Let us consider the 1 light-cone limit with $\bl\ll1$  while  
\be
\Tau=\ln\frac T{\al a}\,,\qquad\Sigma=\ln\frac Sa
\label{TauSigma}
\ee
are both large for the double Logs to be of order 1. 
We assume  $\al>0$ 
(for $\al<0$ it should be substituted by $|\al|$). 

The double Logs appear in the ladder equation 
\rf{nonladder2} from the domain of integration $t\gg \al s$, 
so we rewrite it with DLA as 
\be
\G\left(S,T;a,b\leq\al a\right)=
1-\bl \int_a^{{\rm min}\left\{S,T/\al \right\}} 
\frac {\d s}s\int_{{\rm max}\{\al s,b\}}^{ T} 
\frac {\d t}{t} \;{\G\left(s,t;a,b\leq\al a\right)} \,.
\label{approxintladder}
\ee
We solve it first for  $\G\left(S,T\geq \al S;a,b\right)$, which
then  determines  $\G\left(S,T\leq \al S;a,b\right)$ in DLA.

\tit{Leading-order solution}

An exact solution to \eq{approxintladder} for $b\leq \al a$
is given by the sum of two Bessel functions
\be
\G\left(S,T\geq \al S;a,b \leq \al a\right)
=J_0\left( 2\sqrt{\bl\Sigma\Tau}\right)+
\frac\Sigma\Tau J_2\left(2\sqrt{\bl\Sigma\Tau}\right).
\label{twoBessels}
\ee
The first term on the RHS of \eq{twoBessels} is familiar from the
$\al=0$ solution. But now the solution is for any
$\al$ with DLA.

The solution \rf{twoBessels} obeys the boundary condition
$\G(a,T;a,b)=1$. The second one, $\G(S,b;a,b)=1$, 
cannot be verified
because $T\geq \al S$. Instead~\rf{twoBessels} obeys 
\be
\frac{\partial}{\partial S} \;\G(S,T\geq \al S;a,b)\Big|_{S=T/\al}
=0
\label{b.c.2}
\ee
at the boundary
$S=T/\al$, which can be deduced from \eq{approxintladder}.

Substituting $S=T/\al$ we have from \eq{twoBessels}
\be 
\G\left(T,T;a,b\leq\al a\right)=
\frac{J_1\left( 2\sqrt{\bl}\Tau\right)}{\sqrt{\bl}\Tau}
\label{J1}
\ee
which is the same Bessel function as in {{Y.~M., Olesen, Semenoff (2006)}}.

Finally, substituting \eq{twoBessels} into \eq{approxintladder}, we get 
\be 
\G\left(S\geq T/\al,T;a,b\leq\al a\right)
=\frac{J_1\left( 2\sqrt{\bl}\Tau\right)}
{\sqrt{\bl}\Tau}
\label{J1m}
\ee
which does not depend on $\Sigma$ with DLA.
We can set $S=\infty$ in \eq{J1m}.

\setcounter{subsection}{0}
\setcounter{equation}{0}
\appendix{Exact Sum of Lightcone Ladders}

I describe in this appendix now to draw the contour in the complex
$\omega$-plane for the solution~\rf{ansatz} with $F$ obeying \eq{hg} 
to satisfy the boundary condition~\rf{bc}
and correspondingly to be a solution for the sum of the lightcone ladder 
diagrams.

\tit{The exact solution}

The ansatz~\rf{ansatz} with $F$ obeying \eq{hg} 
will satisfy the boundary condition~\rf{bc} if the integrand has no poles  
in the complex $\omega$-plane. Then the contour of the integration
over $\omega$ can be arbitrarily deformed.

The following linear combination of solutions of the hypergeometric 
equation~\rf{hg} solves the problem as was shown by 
{Y.~M., Olesen, Semenoff (2006)}: 
\begin{eqnarray}
\lefteqn{G (S,T;a,b)=\oint\limits_{C^r} \frac{\d\omega}{2\pi i\omega}~
{}_2 F_1\left(-\sqrt{\bl}\omega,
-\sqrt{\bl}\omega^{-1};
1-\sqrt{\bl}(\omega+\omega^{-1});-\al\frac ab \right)}
\nonumber \\* && \hspace*{1cm}
\times\left( \frac Sa \right)^{\sqrt{\bl}\omega }
\left(\frac Tb \right)^{-\sqrt{\bl}\omega^{-1} }
{}_2 F_1 \left( \sqrt{\bl}\omega ,\sqrt{\bl}\omega^{-1};
1+\sqrt{\bl}(\omega +\omega^{-1});
-\al \frac ST  \right) \nonumber \\
&&+ \int_{|_ {n_{\rm min}-0}} \frac {d s}{2\pi i}
\frac{\Gamma(-s)}{\Gamma(s)}
\frac{1}{\sqrt{\bl}(\omega_+^R-\omega_-^R)}
\frac{\Gamma(\sqrt{\bl}\omega_+^R)\Gamma(1+\sqrt{\bl}\omega_+^R)}
{\Gamma(\sqrt{\bl}\omega_+^L)\Gamma(1+\sqrt{\bl}\omega_+^L)}
\nonumber \\* && \hspace{1cm} \times 
\left[\left(\frac {\al S}b\right)^{\sqrt{\bl}\omega_+^R}\,
\left(\frac T{\al a}\right)^{-\sqrt{\bl}\omega_-^R} +
 \left(\frac {\al S}b\right)^{\sqrt{\bl}\omega_-^R}\,
\left(\frac T{\al a} \right)^{-\sqrt{\bl}\omega_+^R}  \right]
\nonumber \\* && \hspace*{1cm} \times
{}_2 F_1\left(\sqrt{\bl}\omega_+^R,
\sqrt{\bl}\omega_-^R; s+1;-\al\frac ab \right)
{}_2 F_1 \left( \sqrt{\bl}\omega_+^R ,\sqrt{\bl}\omega_-^R;
s+1; -\al \frac ST  \right) \nonumber \\*
&&
\label{aaremarkable}
\end{eqnarray}
with
\be
\omega^R_\pm  = \frac s{2\sqrt{\bl}} \pm \sqrt{\frac{s^2}{4\bl}-1}
\qquad
\omega^L_\pm  =  -\omega_\mp^R=
-\frac s{2\sqrt{\bl}} \pm \sqrt{\frac{s^2}{4\bl}-1} \,.
\ee

The contour
$C^r$ in the first term  runs over a circle of arbitrary radius $r$
($|\omega|=r$). The second contour integral runs parallel to
imaginary axis along 
\be
s=n_{\rm min}-0 +ip\qquad(-\infty<p<+\infty) \,,
\ee
where
\begin{equation}
n_{\rm min}=\left[\sqrt{\bl}(r+1/r)\right]+1
\label{nmin}
\end{equation}
and $[\cdots]$ denotes the integer part.

\tit{Cancellation of poles}

\begin{figure}[h]
\vspace*{5mm}
\centering
{\includegraphics[width=10cm]{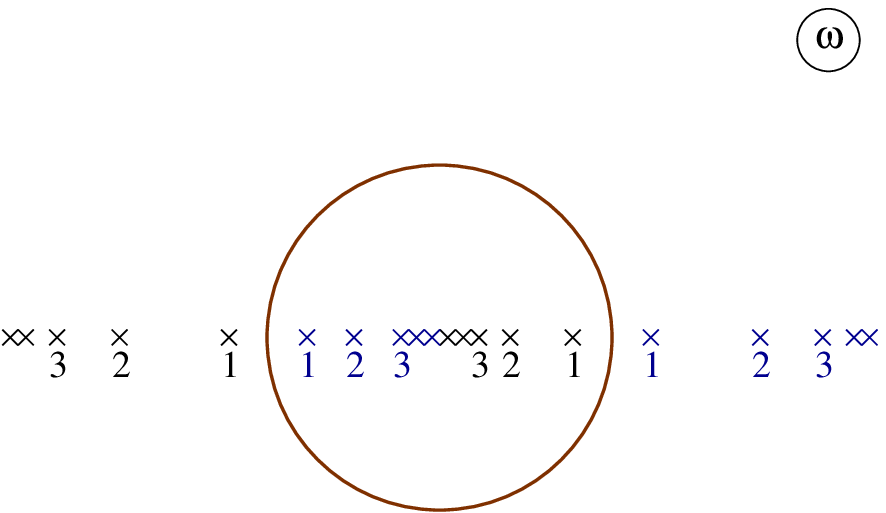}
}
\centering
{\includegraphics[width=10cm]{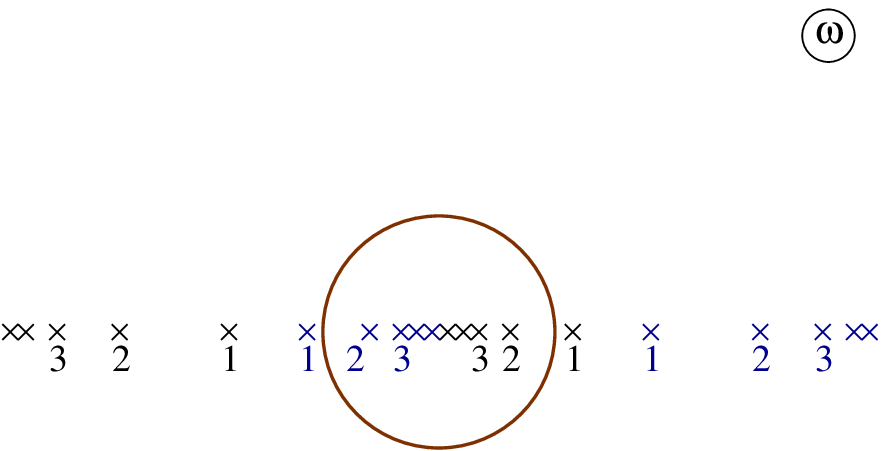}
}\caption[location of poles]  
{Location of poles of the integrand 
in the first term in \eq{aaremarkable} for $ \bl<\frac14$, $\tbr{r}=1$
(above) and $ \bl<\frac14$, $\tbr{r}<1$ (below).
The circle represents the contour of integration $C^r$.}
\label{fi:poles}
\end{figure}

The integrand in the first term on the right-hand side of \eq{aaremarkable}
has poles at 
\be
\omega=\omega_\pm^R(n)=\frac n{2\sqrt{\bl}} \pm \sqrt{\frac{n^2}{4\bl}-1} \,,
\qquad \omega=\omega_\pm^L(n)=
-\frac n{2\sqrt{\bl}} \pm \sqrt{\frac{n^2}{4\bl}-1}
\ee
as the hypergeometric functions have. These poles are depicted in 
Fig.~\ref{fi:poles}.

The poles of the integrand in the first term
are cancelled by the poles at $s=n$ of the integrand in the second term,
so the contour of integration can be arbitrarily deformed.
This is guaranteed by the value of $n_{\rm min}$, given by \eq{nmin}, which
changes accordingly with $r$ to provide the cancellation of the poles
of the first term, that are located inside $C^r$, by those 
of the second term, that are located to the right on $n_{\rm min}$.

The boundary conditions~\rf{bc} are satisfied by the
solution~\rf{aaremarkable}: we choose the integration contour in
the first term to be a circle of the radius which is
either  small for $T=b$ or  large for $S=a$.
Then 
the residue at the pole at $\omega=0$ or $\omega=\infty$
equals 1 which proves that the boundary conditions is satisfied.

For $S=T$, $a=b$ and $\al=-1$ the solution~\rf{aaremarkable} simplifies to 
the Bessel function~\rf{GfinJ}.



\begin{thebibliography}{99}
\addtolength{\itemsep}{-2pt}

\bibitem{AM07a}
  L.~F.~Alday and J.~Maldacena,
 {\it Gluon scattering amplitudes at strong coupling,}
  JHEP {\bf 0706} (2007) 064 
  [arXiv:0705.0303 [hep-th]].

\bibitem{AM07b}
L.~F.~Alday and J.~Maldacena,
{\it Comments on operators with large spin,} JHEP {\bf 0711} (2007) 019. 
  [arXiv:0708.0672 [hep-th]].

\bibitem{AM07c}
  L.~F.~Alday and J.~Maldacena,
  {\it Comments on gluon scattering amplitudes via AdS/CFT,}
  JHEP {\bf 0711} (2007) 068
  [arXiv:0710.1060 [hep-th]].

\bibitem{BKK07}
B.~Basso, G.~P.~Korchemsky and J.~Kotanski,
{\it Cusp anomalous dimension in maximally supersymmetric Yang-Mills theory at
 strong coupling,} Phys.\ Rev.\ Lett.\ {\bf 100} (2008) 091601 
  [arXiv:0708.3933 [hep-th]].

\bibitem{BES07}
N.~Beisert, B.~Eden and M.~Staudacher,
{\it Transcendentality and crossing,}
  J.\ Stat.\ Mech.\  {\bf 0701} (2007) P021
  [arXiv:hep-th/0610251].

\bibitem{BBKS}
M.~K.~Benna, S.~Benvenuti, I.~R.~Klebanov and A.~Scardicchio,
{\it A test of the AdS/CFT correspondence using high-spin operators,}
Phys.\ Rev.\ Lett.\  {\bf 98}, 131603 (2007)
  [{arXiv:hep-th/0611135}].

\bibitem{BCFM98}
D.~Berenstein, R.~Corrado, W.~Fischler and J.~Maldacena,
{\it The operator product expansion for Wilson loops and surfaces in the
large $N$ limit,} Phys.\ Rev.\ D {\bf 59} (1999) 105023
[{arXiv:hep-th/9809188}].

\bibitem{BDS05}
Z.~Bern, L.~J.~Dixon and V.~A.~Smirnov,
  {\it Iteration of planar amplitudes in maximally supersymmetric Yang-Mills
  theory at three loops and beyond,}
  Phys.\ Rev.\ D {\bf 72} (2005) 085001
  [{arXiv:hep-th/0505205}].

\bibitem{BCDKS}
 Z.~Bern, M.~Czakon, L.~J.~Dixon, D.~A.~Kosower and V.~A.~Smirnov,
 {\it The four-loop planar amplitude and cusp anomalous dimension in maximally
 supersymmetric Yang-Mills theory,}
  Phys.\ Rev.\  D {\bf 75} (2007) 085010
  [arXiv:hep-th/0610248].

\bibitem{BNS81}
{R.~A. Brandt, F. Neri  and M.~Sato,}
{\it Renormalization of loop functions for all loops},
Phys.\ Rev.\ D {\bf 24} (1981) 879.


\bibitem{Bro80}
S.~J.~Brodsky, Y.~Frishman, G.~P.~Lepage, and C.~T.~Sachrajda,
{\it Hadronic wave functions at short distances and the operator product 
expansion,}
Phys.\ Lett.\ B {\bf 91} (1980) 239.

\bibitem{DV80}
{V.~S. Dotsenko and S.~N. Vergeles,}
{\it Renormalizability of phase factors in non-Abelian gauge theory},
Nucl.\ Phys.\ B {\bf 169} (1980) 527.

\bibitem{DG00}
N.~Drukker and D.~J.~Gross, {\it An exact prediction of ${\cal N} =
4$ SUSYM theory for string theory,} J.\ Math.\ Phys.\  {\bf 42}
(2001) 2896 [{arXiv:hep-th/0010274}].

\bibitem{DGO99}
N.~Drukker, D.~J.~Gross and H.~Ooguri,
{\it Wilson loops and minimal surfaces},
Phys.\ Rev.\ D {\bf 60} (1999) 125006
[{arXiv:hep-th/9904191}].

\bibitem{DHKS}
J.~M.~Drummond, J.~Henn, G.~P.~Korchemsky and E.~Sokatchev,
{\it On planar gluon amplitudes/Wilson loops duality,}
Nucl.\ Phys.\ B {\bf 795} (2008) 52 [arXiv:0709.2368 [hep-th]].

\bibitem{Drummond:2008vq}
J.~M.~Drummond, J.~Henn, G.~P.~Korchemsky and E.~Sokatchev,
{\it Dual superconformal symmetry of scattering amplitudes in N=4
  super-Yang-Mills theory,}
  arXiv:0807.1095 [hep-th].

\bibitem{DSK07}
J.~M.~Drummond, G.~P.~Korchemsky and E.~Sokatchev,
{\it Conformal properties of four-gluon planar amplitudes and Wilson loops,}
 Nucl.\ Phys.\ B {\bf 795} (2008) 385 [arXiv:0707.0243 [hep-th]].

\bibitem{ES06}
B.~Eden and M.~Staudacher, 
{\it Integrability and transcendentality}, 
J.\ Stat.\ Mech.\ {\bf 0611} (2006) P014
[arXiv:hep-th/0603157].

\bibitem{ESSZ99}
J.~K.~Erickson, G.~W.~Semenoff, R.~J.~Szabo, and K.~Zarembo, {\it
Static potential in ${\cal N} = 4$ supersymmetric Yang-Mills
theory,} Phys.\ Rev.\ D {\bf 61} (2000) 105006 [{arXiv:hep-th/9911088}].

\bibitem{ESZ00}
J.~K.~Erickson, G.~W.~Semenoff and K.~Zarembo,
{\it Wilson loops in ${\cal N} = 4$ supersymmetric Yang-Mills theory,}
Nucl.\ Phys.\ B {\bf 582} (2000) 155
[{arXiv:hep-th/0003055}].

\bibitem{FKKT97}
  M.~Fukuma, H.~Kawai, Y.~Kitazawa and A.~Tsuchiya,
  {\it String field theory from IIB matrix model,}
  Nucl.\ Phys.\ B {\bf 510} (1998) 158
  [{arXiv:hep-th/9705128}].

\bibitem{FT02}
S.~Frolov and A.~A.~Tseytlin
{\it Semiclassical quantization of rotating superstring in $AdS_5 \times S^5$,}
JHEP {\bf 0206} (2002) 007 [{arXiv:hep-th/0204226}].

\bibitem{HM89}
{M.~B.~Halpern and Y.~M.~Makeenko,}
{\it Continuum-regularized loop-space equation},
Phys.\ Lett.\ B {\bf 218} (1989) 230.

\bibitem{GN80a}
J.~L. Gervais and A. Neveu, {\it The slope of the leading Regge 
trajectory in quantum chromodynamics}, 
Nucl.\ Phys.\ B {\bf 163} (1980) 189.

\bibitem{GW73}
D.~J.~Gross and F.~Wilczek,
{\it Asymptotically free gauge theories,}
Phys.\ Rev.\  D {\bf 8} (1973) 3633.

\bibitem{GKP02}
S.~S.~Gubser, I.~R.~Klebanov and A.~M.~Polyakov,
{\it A semi-classical limit of the gauge/string correspondence},
Nucl.\ Phys.\  B {\bf 636} (2002) 99
[{arXiv:hep-th/020405}].

\bibitem{KM93}
G.~P.~Korchemsky and G.~Marchesini, {\it Partonic distributions for large
$x$ and renormalization of Wilson loop}, Nucl.\ Phys.\  B {\bf 406} (1993) 225.

\bibitem{KR87}
G.~P.~Korchemsky and A.~V.~Radyushkin, {\it Renormalization of the Wilson
loops beyond the leading order},
Nucl.\ Phys.\  B {\bf 283} (1987) 342.

\bibitem{KL06}
   A.~V.~Kotikov and L.~N.~Lipatov,
   {\it On the highest transcendentality in N = 4 SUSY,}
   Nucl.\ Phys.\  B {\bf 769} (2007) 217
   [arXiv:hep-th/0611204].

\bibitem{KLOV}
A.~V.~Kotikov, L.~N.~Lipatov, A.~I.~Onishchenko, and V.~N.~Velizhanin,
{\it Three-loop universal anomalous dimension of the Wilson operators
in ${\cal N} = 4$ SUSY Yang-Mills model,}
Phys.\ Lett.\ B {\bf 595} (2004) 521
[{arXiv:hep-th/0404092}].

\bibitem{Kru02}
M.~Kruczenski,
{\it A note on twist two operators in ${\cal N}=4$ SYM and Wilson
loops in Minkowski signature}, JHEP {\bf 0212} (2002) 024
[{arXiv:hep-th/0210115}].

\bibitem{Mal98b}
J.~Maldacena,
{\it Wilson loops in large $N$ field theories},
Phys.\ Rev.\ Lett.\  {\bf 80} (1998) 4859
[{arXiv:hep-th/9803002}].

\bibitem{Mak81}
Y.~M.~Makeenko,
{\it On conformal operators in quantum chromodynamics,}
Sov.\ J.\ Nucl.\ Phys.\  {\bf 33} (1981) 440.

\bibitem{Mak88}
Y.~M.~Makeenko,
  {\it Polygon discretization of the loop space equation,}
  Phys.\ Lett.\ B {\bf 212} (1988) 221.

\bibitem{Mak02}
Y.~Makeenko, {\it Methods of contemporary gauge theory},
Cambridge Univ.\ Press (2002). 

\bibitem{Mak03}
Y.~Makeenko,
{\it Light-cone Wilson loops and the string/gauge correspondence,}
JHEP {\bf 0301} (2003) 007
[{arXiv:hep-th/0210256}].

\bibitem{MM79}
  Y.~M.~Makeenko and A.~A.~Migdal,
  {\it Exact equation for the loop average in multicolor QCD,}
  Phys.\ Lett.\ B {\bf 88} (1979) 135.

\bibitem{MM81}
Y.~M.~Makeenko and A.~A.~Migdal,
{\it Quantum chromodynamics as dynamics of loops,}
Nucl.\ Phys.\ B {\bf 188} (1981) 269.

\bibitem{MOS}
Y.~Makeenko, P.~Olesen and G.~W.~Semenoff,
{\it Cusped SYM Wilson loop at two loops and beyond},
Nucl.\ Phys.\ B {\bf 748} (2006) 170 [{arXiv:hep-th/0602100}].

\bibitem{MVV04}
S.~Moch, J.~A.~M.~Vermaseren and A.~Vogt,
{\it The three-loop splitting functions in QCD: The non-singlet case,}
Nucl.\ Phys.\  B {\bf 688} (2004) 101
[arXiv:hep-ph/0403192].

\bibitem{Ond82}
T.~Ohrndorf,
{\it Constraints from conformal covariance on the mixing of operators 
of lowest twist,}
Nucl.\ Phys.\ B {\bf 198} (1982) 26.

\bibitem{Pol80}
{A.~M.~Polyakov,}
{\it Gauge fields as rings of glue},
Nucl.\ Phys.\ B {\bf 164} (1980) 171.

\bibitem{RY98}
S.-J.~Rey and J.~Yee,
{\it Macroscopic strings as heavy quarks in large $N$ gauge theory and
anti-de Sitter supergravity,}
Eur.\ Phys.\ J.\ C {\bf 22} (2001) 379
[{arXiv:hep-th/9803001}].

\bibitem{RT07}
R.~Roiban and A.~A.~Tseytlin,
{\it Strong-coupling expansion of cusp anomaly from quantum superstring,}
JHEP {\bf 0711} (2007) 016 [arXiv:0709.0681 [hep-th]].

\bibitem{RT08}
R.~Roiban and A.~A.~Tseytlin,
{\it Spinning superstrings at two loops: strong-coupling corrections to
  dimensions of large-twist SYM operators,}
  Phys.\ Rev.\  D {\bf 77} (2008) 066006
  [arXiv:0712.2479 [hep-th]].

\bibitem{SZ02}
G.~W.~Semenoff and K.~Zarembo,
{\it Wilson loops in SYM theory: from weak to strong coupling,}
Nucl.\ Phys.\ Proc.\ Suppl.\ {\bf 108} (2002) 106 [{arXiv:hep-th/0202156}].

\bibitem{Sta05}
M.~Staudacher, {\it The factorized S-matrix of CFT/AdS},
JHEP {\bf 0505} (2005) 054
[{arXiv:hep-th/0412188}].

\bibitem{Wil74}
{K.~G. Wilson,} {\it Confinement of quarks}, 
Phys.\ Rev.\ D {\bf 10} (1974) 2445.

\end{thebibliography}
\end{document}